# Modeling of Nucleation Processes

**Emmanuel Clouet**
CEA, DEN, Service de Recherches de Métallurgie Physique, France
✉ emmanuel.clouet@cea.fr

Nucleation is the onset of a first-order phase transition by which a metastable phase transforms into a more stable one. Such a phase transition occurs when an initial system initially in equilibrium is destabilized by the change of an external parameter like the temperature or the pressure. If the perturbation is small enough, the system does not become unstable but rather stays metastable. In diffusive transformations, the system then evolves through the nucleation, the growth and the coarsening of a second phase. Such a phase transformation is found in a lot of situations in materials science like condensation of liquid droplets from a supersaturated vapor, solidification, precipitation from a supersaturated solid solution, … The initial stage of all these different processes can be well described within the same framework currently know as the classical nucleation theory.

Since its initial formulation in 1927 by Volmer, Weber and Farkas [1, 2] and its modification in 1935 by Becker and Döring [3] the classical nucleation theory has been a suitable tool to model the nucleation stage in phase transformations. The success of this theory relies on its simplicity and on the few parameters required to predict the nucleation rate, *i.e.* the number of clusters of the new phase appearing per unit of time and of volume. It allows rationalizing experimental measurements, predicting the consequences of a change of the control parameters like the temperature or the supersaturation, and describing nucleation stage in mesoscopic modeling of phase transformations.

In this article, we first describe the results obtained by Volmer, Weber, Farkas, Becker and Döring [1, 2, 3] and which constitutes the classical nucleation theory. These results are the predictions of the precipitate size distribution, of the steady-state nucleation rate and of the incubation time. This theory describes the nucleating system as a homogeneous phase where heterophase fluctuations occur. Some of these fluctuations reach a size big enough so that they can continue to grow and lead to the formation of precipitates. The nucleating system is thus envisioned mainly from a thermodynamic view-point. The key controlling parameters are the nucleation driving force and the interface free energy. A kinetic approach, the cluster dynamics, can also be used to describe nucleation. This constitutes the second part of this article. Here, a master equation describes the time evolution of the system which is modeled as a cluster gas. The key parameters are the cluster condensation and evaporation rates. Both approaches are different in their description of the nucleating system and in their needed input parameters. They are nevertheless closely related. Predictions of the classical nucleation theory have been actually derived from the same master equation used by cluster dynamics [3], and extensions of classical nucleation theory always starts from this master equation. In this article the links as well as the difference between both descriptions are emphasized. Since its initial formulation, the classical nucleation theory has been enriched, mainly by Binder and Stauffer [4, 5, 6], so as to take into account the fact that clusters other than monomers can migrate and react. It has been also extended to multi-component systems [7, 8, 9, 10, 11, 12]. These generalizations of the initial formalism are presented at the end of the second part.

# Thermodynamic Approach

### Conditions for Nucleation

Nucleation occurs when a homogeneous phase initially in stable thermal equilibrium is put in a state where it becomes metastable by the variation of a controlling parameter. In the following case, the controlling parameter is the temperature and the initial system is quenched through a first-order phase transition in a two-phase region. The system then tends to evolve toward a more stable state and to reach its equilibrium. As the parent phase is not unstable, this transformation cannot proceed through the continuous development of growing infinitesimal perturbations delocalized in the whole phase, *i.e.* by spinodal decomposition [13, 14]. Such perturbations in a metastable state increase the free energy. As a consequence, they can appear because of thermal fluctuations but they naturally decay. To reach its equilibrium, the system has to overcome an energy barrier so as to form directly clusters of the new equilibrium phase, a process known as nucleation.

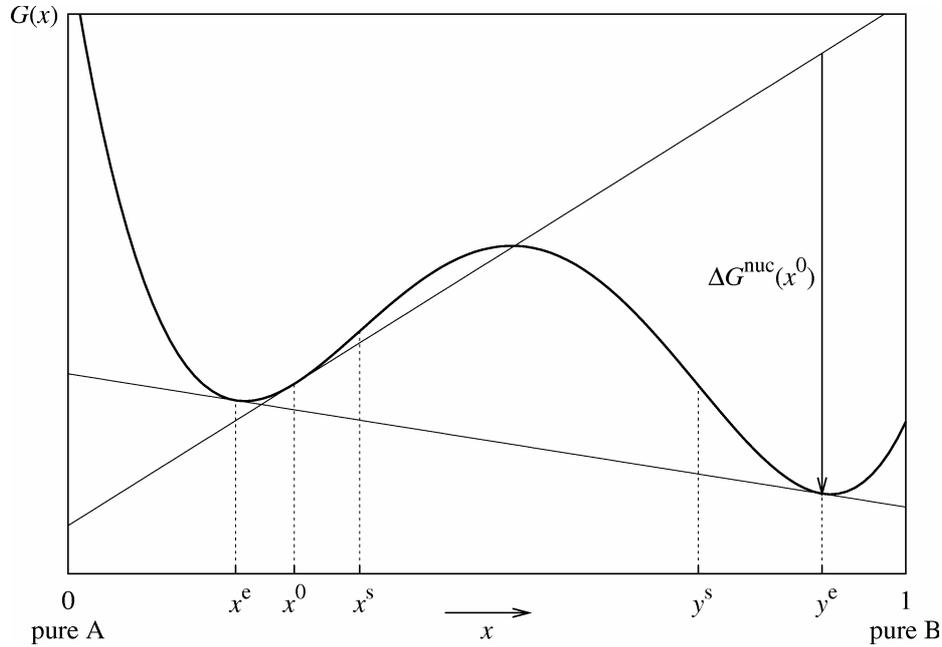

**Figure 1:** Sketch of the free energy of a binary mixture quenched in a two-phase region. The bold line is the free energy per atom $G(x)$ of the homogeneous system. The compositions $x^e$ and $y^e$ of the equilibrium phases are given by the common tangent construction. The spinodal limits $x^s$ and $y^s$ define the unstable region. $\Delta G^{nuc}(x^0)$ is the nucleation free energy of the metastable homogeneous system of composition $x^0$.

This difference between a metastable and an unstable state, as well as between nucleation and spinodal decomposition, is better understood through the following example. We consider a system corresponding to a binary mixture of two elements A and B with a fixed atomic fraction $x$ of B elements. Such a system can be a solid or a liquid solution for instance. We assume that the free energy per atom $G(x)$ of this system is known for every composition $x$ and is given by the function plotted in Fig. 1. A two-phase region given by the common tangent construction exists at the considered temperature: the equilibrium state of binary mixtures with an intermediate composition $x^0$ between $x^e$ and $y^e$ corresponds to a mixture of two phases having the compositions $x^e$ and $y^e$. A homogeneous system with a composition $x^0$ will then separate into these two equilibrium phases. We can examine the variation of the free energy if this transformation happens through the development of infinitesimal fluctuations. In that purpose, we consider a small perturbation corresponding to a separation of the initially homogeneous system into two phases having the compositions $x^0 + dx_1$ and $x^0 + dx_2$. As we want the perturbation to be small, we assume $|dx_1| \ll 1$ and $|dx_2| \ll 1$. If $f_1$ is the fraction of phase 1, matter conservation imposes the following relation between both compositions

$$f_1 dx_1 + (1-f_1) dx_2 = 0. \tag{1}$$

The free energy variation associated with this unmixing is given by

$$\begin{aligned}\Delta G &= f_1 G(x^0 + dx_1) + (1-f_1) G(x^0 + dx_2) - G(x^0) \\ &= \frac{1}{2}\left[f_1 dx_1^2 + (1-f_1) dx_2^2\right] G''(x^0) + o(dx_1^2)\end{aligned} \tag{2}$$

The first derivative $G'(x^0)$ of the free energy does not appear in equation (2) because of the relation (1). The sign of the free energy variation is thus governed by the second derivative $G''(x^0)$ of the free energy. If this second derivative is negative, the initial infinitesimal perturbation decreases the free energy (Fig. 2a). It can therefore develop until the system reaches its two phase equilibrium state. This is the regime of spinodal decomposition. In Fig. 1, the free energy second derivative changes its signs in $x^s$ and $y^s$: all homogeneous systems with a composition between these limits are unstable and evolve spontaneously to equilibrium. On the other hand, if the composition $x^0$ is higher than the equilibrium composition $x^e$ but smaller than the spinodal limit $x^s$, the homogeneous binary mixture is metastable. As the second derivative of the free energy is positive, any infinitesimal perturbation increases the free energy (Fig. 2b) and will therefore decay. To reach its equilibrium state, the system has to overcome an energy barrier and the phase separation occurs by nucleation of the new equilibrium phase with the composition $y^e$.

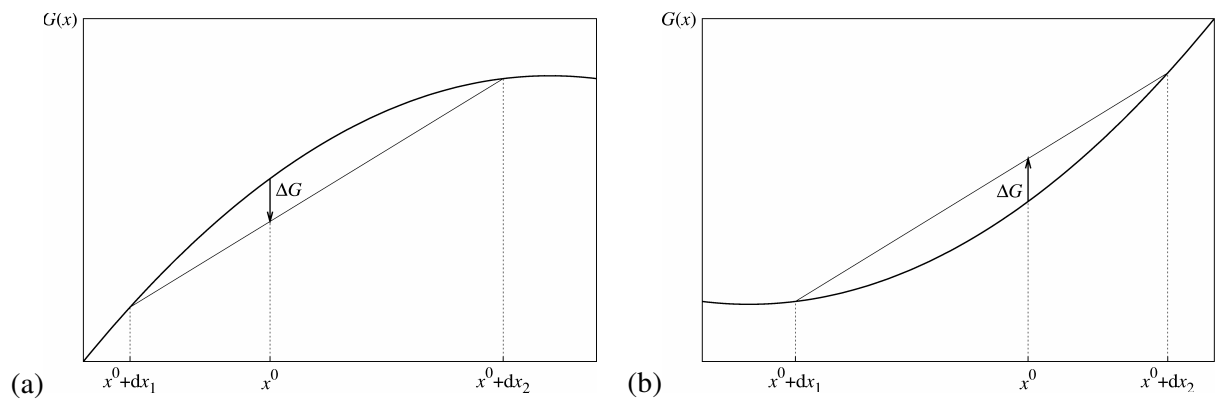

**Figure 2**: Variation $\Delta G$ of the free energy corresponding to the spontaneous unmixing of a homogeneous system of composition $x^0$ in two phases of respective compositions $x^0 + dx_1$ and $x^0 + dx_2$. (a): spinodal regime ($G''(x^0) < 0$); (b): nucleation regime ($G''(x^0) > 0$).

### *The Capillary Approximation*

In the nucleation regime, the system evolves through the formation of well-defined and localized fluctuations corresponding to clusters of the new equilibrium phase. The formation free energies of these clusters are well described by the capillary approximation. This assumes that two contributions enter this free energy (Fig. 3):

- Volume contribution: by forming a cluster of the new phase, the system decreases its free energy. The gain is directly proportional to the volume of the cluster or, equivalently, to the number $n$ of atoms forming the cluster. This is the nucleation driving force.
- Surface contribution: one needs to create an interface between the parent phase and the cluster of the new phase. This interface has a cost which is proportional to the surface area of the cluster or, equivalently, to $n^{(d-1)/d}$ where $d$ is the dimension of the system.

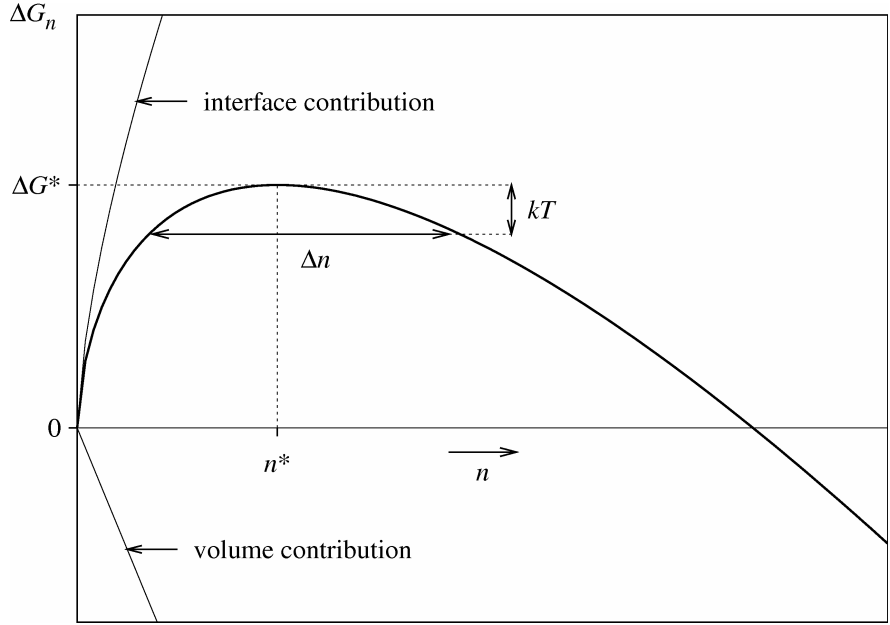

**Figure 3:** Variation of the cluster formation free energy $\Delta G_n$ with the number $n$ of atoms they contain as described by Eq. (3). $n^*$ is the critical size and $\Delta G^*$ the corresponding free energy. The size interval $\Delta n$ characterizes the energy profile around the critical size and is directly linked to the Zeldovich factor (Eq. (10)).

In the following, we restrict ourselves to the 3 dimension case. The formation free energy of a cluster containing $n$ atoms is then given by

$$\Delta G_n = n\Delta G^{\text{nuc}} + n^{2/3} A\sigma, \qquad (3)$$

where $\Delta G^{\text{nuc}}$ is the nucleation free energy, $\sigma$ the interface free energy and $A$ a geometric factor. If the interface free energy is isotropic, the equilibrium shape of the cluster is a sphere. The corresponding geometric factor is then $A = \left(36\pi\Omega_1^2\right)^{1/3}$, where $\Omega_1$ is the volume of a monomer. For anisotropic interface free energy, one can use the Wulff construction [15, 16] to determine the equilibrium shape, *i.e.* the shape with minimum free energy for a given volume, and deduce an average interface free energy corresponding to a hypothetical spherical cluster having the same volume and the same interface energy as the real one which may be facetted. An example is given in Ref. 17 for precipitates with {100}, {110}, and {111} interfaces.

The nucleation free energy is obtained by considering the difference of chemical potentials in the parent and in the equilibrium phases for all atoms composing the cluster,

$$\Delta G^{\text{nuc}} = \sum_i y_i^{\text{e}} (\mu_i^{\text{e}} - \mu_i^0), \qquad (4)$$

where $y_i^{\text{e}}$ is the atomic fraction of the type $i$ atom in the nucleating equilibrium phase, $\mu_i^{\text{e}}$ and $\mu_i^0$ are the corresponding chemical potentials respectively in the nucleating equilibrium phase and in the parent phase. When the parent phase is metastable, chemical potentials in this phase are higher than the ones at equilibrium. The nucleation free energy given by Eq. (4) is therefore negative. Classic expressions of the nucleation free energy are given at the end of this section in some simple cases.

For negative nucleation driving force, because of the competition between the volume and the interface contributions, the cluster formation free energy (Eq. (3)) shows a maximum for a given critical size $n^*$ as illustrated in Fig. 3. $n^*$ corresponds to the size at which the first derivative of $\Delta G_n$ is equal to zero, thus leading to

$$n^* = \left(-\frac{2}{3}\frac{A\sigma}{\Delta G^{\text{nuc}}}\right)^3, \qquad (5)$$

and the corresponding formation free energy

$$\Delta G^* = \Delta G_{n^*} = \frac{4}{27}\frac{(A\sigma)^3}{(\Delta G^{nuc})^2}. \tag{6}$$

Below this critical size, the energy of growing clusters increases because of the interface predominance at small sizes. Clusters in this range of size are therefore unstable: if a cluster is formed, it will tend to redissolve. Unstable clusters can be nevertheless found in the parent phase because of thermal fluctuations. The size distribution of these clusters is given by

$$C_n^{eq} = C_0 \exp\left(-\frac{\Delta G_n}{kT}\right), \tag{7}$$

where $C_0$ is the atomic fraction of sites accessible to the clusters. For precipitation in the solid state for instance, all lattice sites can receive a cluster and therefore $C_0 = 1$. The validity of the size distribution (7) can be demonstrated for an undersaturated system ($\Delta G^{nuc} \geq 0$) using a lattice gas model (compare with Cluster Gas Thermodynamics in the Kinetic Approach section of this article). For a supersaturated system, one assumes that the system reaches a steady-state where clusters smaller than the critical size still obey the distribution (7).

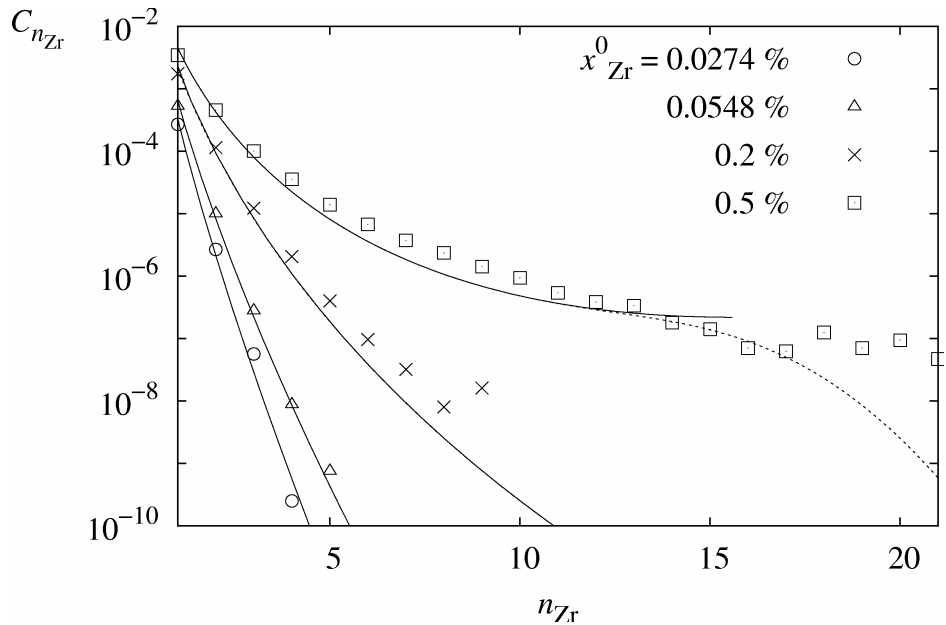

**Figure 4:** Dependence on the nominal concentration $x_{Zr}^0$ of the cluster size distribution of an aluminium solid solution at 500°C. At this temperature, the solubility limit is $x_{Zr}^e = 0.0548\,\%$. Symbols correspond to atomic simulations (kinetic Monte Carlo) [18] and lines to predictions of the classical nucleation theory: full lines to the equilibrium cluster size distribution (Eq. (7)) for $n \leq n^*$ and dotted lines to the steady-state distributions (Eqs. (8) and (85)).

Comparisons with atomic simulations have shown that Eq. (7) correctly describes the size distribution of sub-critical clusters. An example of such a comparison is given in Fig. 4 for Al-Zr alloys leading to the coherent precipitation of $L1_2$ $Al_3Zr$ compounds [17]: size distributions are given for undersaturated, saturated, and supersaturated solid solutions. A similar comparison leading to the same conclusion can be found in Ref. 18 for an unmixing alloy on a body centered cubic lattice, or in Ref. 19 and 20 for the magnetization reversal of an Ising model respectively in 2 and 3 dimensions.
The kinetic approach developed further in this article shows that the steady-state distribution in a nucleating system actually slightly deviates from the equilibrium distribution (7) around the critical size. An exact expression of the steady-state distribution has been obtained by Kashiev [21, 22]. In the critical size interval $\Delta n$, which will be precisely defined below, it can be approximated by

$$C_n^{st} = \left[\frac{1}{2} - Z(n - n^*)\right] C_n^{eq}. \tag{8}$$

The Zeldovich factor appearing in this equation is a function of the second derivative of the cluster formation free energy at the critical size,

$$Z = \sqrt{-\frac{1}{2\pi kT} \frac{\partial^2 \Delta G_n}{\partial n^2}\bigg|_{n=n^*}} = \frac{3(\Delta G^{nuc})^2}{4\sqrt{\pi kT}(A\sigma)^{3/2}}. \tag{9}$$

The physical meaning of the Zeldovich factor can be seen in Fig. 3 which sketches the variation of the cluster formation free energy with their size. If the formation free energy was harmonic, the size interval where the difference between the cluster free energy and the nucleation barrier $\Delta G^*$ is smaller than the thermal energy $kT$ would be given by

$$\Delta n = \frac{2}{\sqrt{\pi}} \frac{1}{Z}. \tag{10}$$

The Zeldovich factor therefore characterizes the flatness of the energy profile around the critical size. Equation (8) shows that steady-state cluster concentrations in the critical region are reduced compared to the equilibrium distribution. For the critical size a factor 1/2 appears in front of the equilibrium concentration.

### *Steady-State Nucleation Rate*

When the nucleation barrier $\Delta G^*$ is high enough compared to the thermal energy $kT$, the metastable state of the system contains thermal fluctuations well described by the distribution (7). Sometimes, one of these fluctuations will reach and overcome the critical size. It can then continue to grow so as to become more and more stable. Classical nucleation theory assumes that the system reaches a steady-state and it then shows that stable nuclei appear at a rate given by [3]

$$J^{st} = \beta^* Z C_0 \exp\left(-\frac{\Delta G^*}{kT}\right), \tag{11}$$

where $\beta^*$ is the rate at which a critical cluster grows and $Z$ is the Zeldovich factor (Eq. (9)). This factor has been introduced by Becker and Döring [3] so as to describe cluster fluctuations around the critical size and, in particular, the probability for a stable nucleus to redissolve. $ZC_0 \exp(-\Delta G^*/kT)$ is therefore the number of critical clusters that reach a size large enough so that they can continuously grow. The initial expression of the nucleation rate derived by Volmer and Weber [1] and by Farkas [2] did not consider this Zeldovich factor and led to an overestimation of the nucleation rate. A small Zeldovich factor corresponds to a flat energy profile around the critical size. Critical clusters experience then size variations which are mainly random and not really driven by their energy decrease. Some of them will redissolve and not fall in the stable region. This explains why the nucleation rate is reduced by the Zeldovich factor. A more rigorous derivation of the nucleation rate where the Zeldovich factor naturally appears will be given in section, The Link with Classical Nucleation Theory.

An expression for the growing rate $\beta^*$ of the critical cluster is needed. If the growth limiting process is the reaction at the interface to attach the atoms on the critical cluster (ballistic regime), $\beta^*$ is then proportional to the cluster area. Assuming that this reaction is controlled by one type of atoms, one obtain the expression [24]

$$\beta^* = 4\pi r^{*2} \frac{\lambda_i \Gamma_i}{\Omega} \frac{x_i^0}{y_i^e}, \tag{12}$$

where $r^*$ is the radius of the critical cluster, $\lambda_i$ the distance corresponding to the atom last jump to become attached to the critical cluster, $\Gamma_i$ the corresponding reaction frequency, and $\Omega$ the volume corresponding to one atomic site. $x_i^0$ and $y_i^e$ are the respective atomic fraction of the jumping atoms in the metastable parent phase and in the stable nucleating phase.

For solid – solid phase transformations, the critical cluster growth is usually controlled by the long-range diffusion of solute atoms. The critical condensation rate is then obtained by solving the classical diffusion problem associated to a growing spherical particle. If only diffusion of one type of atoms limits the growth and all other atomic species diffuse sufficiently fast enough so that the cluster composition instantaneously adjusts itself, one obtains [24]

$$\beta^* = 4\pi r^* \frac{D_i}{\Omega} \frac{x_i^0}{y_i^e}, \qquad (13)$$

where $D_i$ is the diffusion coefficient of type $i$ atoms. In a multi-component alloy, when diffusion coefficients of different atomic species have close values and when the composition of the critical cluster can vary, one has to use the "linked flux analysis" presented in this article in the section on Cluster Dynamics. In all cases, the growth rate is proportional to the cluster radius in this diffusive regime.

Both events, *i.e.* the long-range diffusion and the reaction at the interface, can be simultaneously taken into account. The corresponding expression of the condensation rate has been derived by Waite [25].

### *Transient Nucleation*

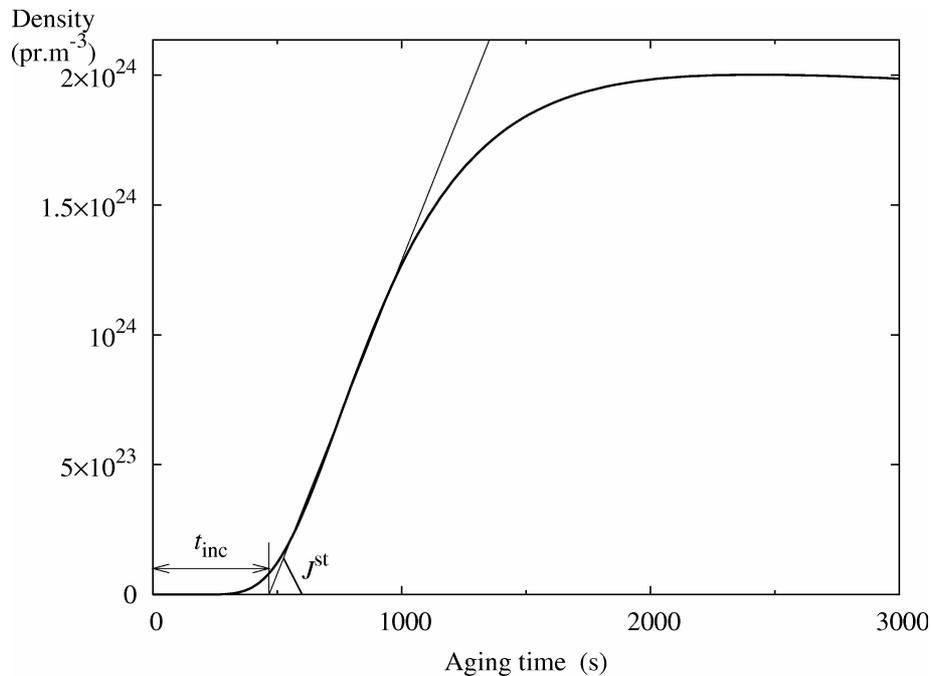

**Figure 5:** Precipitate density as a function of the aging time for an aluminium solid solution containing 0.18 at.% of scandium aged at 300°C. The time evolution obtained from cluster dynamics simulations [23] allows to define a steady-state nucleation rate $J^{st}$ and an incubation time $t_{inc}$.

A transient regime exists before the nucleation rate reaches its stationary value (Eq. (11)). One defines conventionally an incubation time, or a time-lag, to characterize this transient regime. This is defined as the intercept with the time axis of the tangent to the curve representing the variations of the nuclei density (Fig. 5). Exact expressions of the incubation time have been obtained as series of the initial and steady-state cluster size distributions [26, 27]. Different approximations have then been made to evaluate this series and obtain closed forms of the incubation time. They all lead to an incubation time which writes

$$t_{\text{inc}}(n^*) = \theta_0 \frac{1}{\pi Z^2 \beta^*} = \theta_0 \frac{16kT(A\sigma)^3}{9(\Delta G^{\text{nuc}})^4 \beta^*}, \tag{14}$$

where the factor $\theta_0$ depends on the chosen approximation and is close to 1 [28]. Some authors obtained a factor $\theta_0$ that depends slightly on the temperature and on the shape of the cluster formation free energy around the critical size [26, 29]. One can stress however that a precise value of this factor is seldom if ever needed. As it will be shown below, the incubation time depends on too many parameters to be known precisely experimentally. Eq. (14) allows describing its main variation when the temperature or the nucleation driving force are changed: this is usually enough to model incubation and nucleation.

This expression (14) of the incubation time can be obtained from simple physical considerations [24, 30]. The steady-state will be reached once the clusters have grown sufficiently far away from the critical size. One can consider that super-critical clusters have a negligible probability to decay when their size becomes greater than $n^* + \frac{1}{2}\Delta n$. $\Delta n$ characterizes the width of the critical region (Fig. 3) and is related to the Zeldovich factor via Eq. (10). As the energy profile is flat in this neighborhood, one can consider that clusters make a random walk in the size space with a constant jump frequency $\beta^*$. Accordingly, the corresponding time needed to diffuse from $n^*$ to $n^* + \frac{1}{2}\Delta n$ is

$$t_{\text{inc}}(n^*) \sim \frac{\Delta n^2}{4\beta^*}. \tag{15}$$

This leads to the expression (14) with a factor $\theta_0 = 1$.

Different approximations of the nucleation rate in this transient regime can also be found in the literature. Kelton *et al.* [28] have compared these approximations with exact results obtained thanks to a numerical integration of the kinetic equations describing nucleation. They concluded that the best suited approximation to describe the transient nucleation rate is the one obtained by Kashchiev [21, 22],

$$J(t) = J^{\text{st}}\left[1 + 2C\sum_{m=1}^{\infty}(-1)^m \exp\left(-\frac{m^2 t}{\tau}\right)\right], \tag{16}$$

where $C = 1$ for a system initially prepared in a state far from its nucleating metastable state. The time constant is given by

$$\tau = \frac{4}{\pi^3}\frac{1}{Z^2 \beta^*}. \tag{17}$$

When $t > \tau$, one can retain only the first term in the sum appearing in (16). Usually, it is even enough to assume that the nucleation rate behaves like the Heaviside step function, *i.e.* that the nucleation rate reaches its stationary value after an incubation time where no nucleation occurs. The incubation time corresponding to Eq. (16) is $\pi^2 \tau / 6$. Therefore, in Kashchiev treatment, the factor in equation (14) is $\theta_0 = 2/3$.

It is worth saying that the incubation time and the associated transient regime depends on the conditions in which the system has been prepared. Relations (14) and (16) implicitly assume that the quench was done form infinite temperature: no cluster around the critical one existed at the initial time. This may not be true. For instance, the system could have been prepared in an equilibrium state corresponding to a slightly higher temperature where it was stable and then quenched in a metastable state. A cluster distribution corresponding to this higher temperature already exists before the beginning of the phase transformation. If the temperature difference of the quench is small, these preexisting clusters will reduce the incubation period. The dependence of the incubation time with the initial conditions has been observed for instance in atomic simulations for an unmixing binary alloy [18]. Starting from a random solid solution corresponding to an infinite temperature preparation, an incubation time is observed before nucleation reaches its steady-state. If the alloy is annealed above its solubility limit before a quench, the incubation stage disappears if the temperature difference of the quench is not too high. Kashchiev considered the effect of this initial cluster distribution on nucleation in the case of a change in pressure [22, 31]. His results can be easily generalized [28]. To do so, we introduce the supersaturation variation

$$\Delta s = \frac{\Delta G^{nuc}(t=0^-)}{kT(t=0^-)} - \frac{\Delta G^{nuc}(t=0^+)}{kT(t=0^+)}, \tag{18}$$

where $t=0^-$ means that thermodynamic quantities are calculated for the initial state in which the system has been prepared, and $t=0^+$ for the state where nucleation occurs. In his derivation, Kashchiev assumed that the interface free energy of the clusters is the same in both stable and metastable states. The constant entering in the expression (16) of the transient nucleation rate is then

$$C = 1 - \frac{\Delta s}{Z}\exp(-n^*\Delta s), \tag{19}$$

and the corresponding incubation time is multiplied by this constant $C$. The supersaturation variation $\Delta s$ is positive, otherwise nucleation would have happened in the initial state in which the system has been prepared. The existence of an initial cluster size distribution therefore always reduces the incubation time. Nevertheless, $C$ rapidly tends to 1 when the thermodynamic states $t=0^-$ and $t=0^+$ become too different.

By definition, the nucleation rate does not depend on the cluster size in the stationary regime. This property is used in equation (11) to calculate the steady-state nucleation rate $J^{st}$ at the critical size. But the time needed for the stationary regime to develop will of course vary with the cluster size. This means that the incubation time depends on the cluster size at which it is measured. The above defined incubation time corresponds to the critical size. But the smallest cluster size that one can detect experimentally may be significantly larger than the critical size. We therefore need to describe the variation with the cluster size of the incubation time. This problem has been solved by Wu [26] and by Shneidman and Weinberg [29] who showed that the incubation time measured at size $n$ is

$$t_{inc}(n) = t_{inc}(n^*) + \frac{1}{2\pi Z^2 \beta^*}\left\{\theta_1 + \ln\left[\sqrt{\pi}Z(n-n^*)\right]\right\}, \tag{20}$$

for $n > n^* + \Delta n$, i.e. a cluster size outside the critical region. The constant $\theta_1$ is 1 in the expression obtained by Wu and $\theta_1 = \gamma/4 + \ln(2)/2$ for Shneidman and Weinberg, where $\gamma \sim 0.5772$ is Euler's constant.

All the above expressions are obtained in the parabolic approximation, i.e. assuming that the cluster formation free energy is well described by its harmonic expansion around the critical size. According to Shneidman and Weinberg [29], this approximation is highly accurate when calculating the steady-state nucleation rate, but its validity is limited for the incubation time. When considering the exact shape of the cluster formation free energy (Eq. (3)), the expression of the incubation time then depends on the model used for the absorption rate. In all cases, the incubation time at the critical size can be written

$$t_{inc}(n^*) = \frac{1}{2\pi Z^2 \beta^*}\left[\frac{\gamma}{2} + \ln(\sqrt{\pi}Zn^*) - \theta_3\right], \tag{21}$$

where the constant $\theta_3$ differs from 0 when the parabolic approximation is not used. In the ballistic regime, when the condensation rate is proportional to the cluster surface like in (12), the authors obtained $\theta_3 = 1 - \ln(3)$. The incubation time measured at size $n$ is then

$$t_{inc}(n) = \frac{1}{2\pi Z^2 \beta^*}\left\{\left(\frac{n}{n^*}\right)^{1/3} + \ln\left[\left(\frac{n}{n^*}\right)^{1/3} - 1\right] + \gamma - 2 + \ln\left[\frac{6\Delta G^*}{kT}\right]\right\}, \quad \forall n > n^* + \frac{1}{2}\Delta n. \tag{22}$$

In the diffusive regime, when the condensation rate is proportional to the cluster radius like in (13), $\theta_3 = 3/2 - \ln(3)$ and

$$t_{inc}(n) = \frac{1}{2\pi Z^2 \beta^*}\left\{\frac{1}{2}\left[\left(\frac{n}{n^*}\right)^{1/3} + 2\right]^2 + \ln\left[\left(\frac{n}{n^*}\right)^{1/3} - 1\right] + \gamma - \frac{7}{2} + \ln\left[\frac{6\Delta G^*}{kT}\right]\right\}, \quad \forall n > n^* + \frac{1}{2}\Delta n. \tag{23}$$

It should be stressed that all these expressions for incubation time have been obtained in the continuous limit valid for large clusters. The expressions should be used only when the nucleation barrier $\Delta G^*$ is high enough for the critical size not being too small.

### Heterogeneous Nucleation

Until now, only homogeneous nucleation has been considered: it was assumed that nuclei can form anywhere in the system. But it may require less energy for the nuclei to form heterogeneously on preferred nucleation sites. These sites can be at the interface with existing impurities or some lattice defects like grain boundaries or dislocations. The classical theory also allows modeling heterogeneous nucleation after some slight modifications. The first modification is that the parameter $C_0$ appearing in the cluster size distribution (7) is now the number of sites where heterogeneous precipitation can take place. We also need to take into account the decrease of the nuclei free energy when they are located at a preferred nucleation site. Such a decrease usually arises from a gain in the interface free energy: it is more favorable for the nuclei to form on an already existing interface as the cost to create the interface between the old and the new phases is reduced.

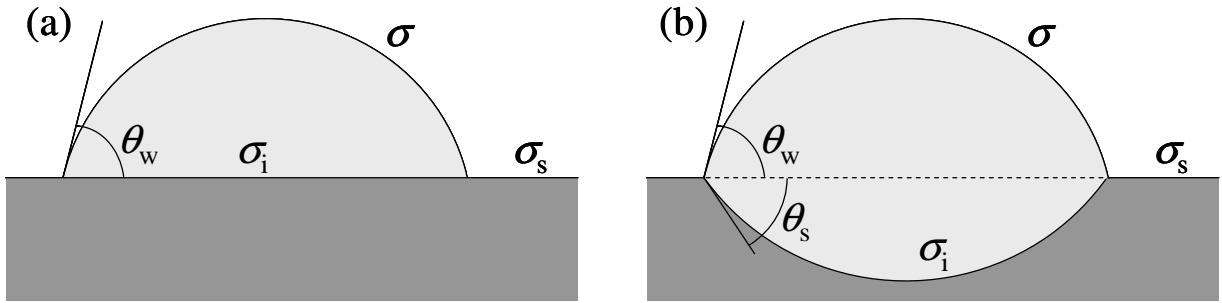

**Figure 6:** Possible shapes of a nucleus in heterogeneous nucleation. (a) Cap shape. (b) Lens shape

First consider the case where the cluster wets the substrate and has a cap shape (Fig. 6a). Electrodeposition is one example where this happens [32]. Three different interface free energies need to be considered:
- $\sigma$ between the parent and the nucleating phase;
- $\sigma_s$ between the parent phase and the substrate;
- $\sigma_i$ between the nucleating phase and the substrate.

The wetting angle is then defined by Young equilibrium equation [16]

$$\cos\theta_w = \frac{\sigma_s - \sigma_i}{\sigma}. \tag{24}$$

The wetting leads to a cap shape only if the interface free energies obey the inequalities $-\sigma \leq \sigma_s - \sigma_i \leq \sigma$. If the difference $\sigma_s - \sigma_i$ is smaller than $-\sigma$, then $\theta_w = \pi$ and the wetting is not possible because an unwet cluster costs less energy. On the other hand, if the difference is greater than $\sigma$, the wetting is complete and one cannot define anymore a cap shape as the nucleating phase will cover uniformly the interface.

The cluster free energy takes the same expression as the one given by the capillary approximation in the homogeneous case (Eq. (3)). To calculate the geometric factor $A$ appearing in this expression, the radius $R$ of the cap must be defined. The cluster volume is then given by [22]

$$n\Omega_1 = \frac{1}{3}\pi R^3 \left(2+\cos\theta_w\right)\left(1-\cos\theta_w\right)^2,$$
$$= \pi R^3 \frac{\left(2\sigma+\sigma_s-\sigma_i\right)\left(\sigma-\sigma_s+\sigma_i\right)^2}{3\sigma^3}, \tag{25}$$

and the free energy associated to the whole cluster interface by

$$n^{2/3} A\sigma = \pi R^2 \left[ \sigma 2(1-\cos\theta_w) + (\sigma_i - \sigma_s)\sin^2\theta_w \right],$$
$$= \pi R^2 (\sigma - \sigma_s + \sigma_i) \frac{2\sigma^2 + (\sigma_i - \sigma_s)(\sigma + \sigma_s - \sigma_i)}{\sigma^2}. \tag{26}$$

Eliminating the variable $R$ between the equations (25) and (26), one obtains the expression of the geometric factor appearing in the capillary approximation

$$A = \left(9\pi\Omega_1^2\right)^{1/3} \frac{2\sigma^2 - (\sigma_s - \sigma_i)(\sigma + \sigma_s - \sigma_i)}{\sigma(2\sigma + \sigma_s - \sigma_i)^{2/3}(\sigma - \sigma_s + \sigma_i)^{1/3}}. \tag{27}$$

When $\sigma_s - \sigma_i = -\sigma$, the unwetting is complete: one recovers the geometric factor $A = \left(36\pi\Omega_1^2\right)^{1/3}$ corresponding to a spherical cluster. With this expression of the geometric factor and the correct value of the parameter $C_0$, all expressions obtained for homogeneous nucleation can also be used for heterogeneous nucleation.

Nuclei can also have a lens shape (Fig. 6b). In such a case, the two wetting angles are defined by [22]

$$\cos\theta_w = \left(\sigma_s^2 + \sigma^2 - \sigma_i^2\right)/2\sigma_s\sigma,$$
$$\cos\theta_s = \left(\sigma_s^2 + \sigma^2 - \sigma_i^2\right)/2\sigma_s\sigma_i. \tag{28}$$

The geometric factor corresponding to this lens shape is obtained using the same method as above: one expresses the volume and the interface energy of the two caps composing the cluster, and one then eliminates the cap radii between these two equations.

### *Examples*

It is worth having a closer look at some examples – solidification and precipitation in the solid state – and to give approximated expression of the nucleation free energy in these simple cases.

#### *Example 1: Solidification*

A single component liquid which was initially at equilibrium is quenched at a temperature $T$ below its melting temperature $T_m$. As the liquid and the solid have the same composition, the nucleation free energy is simply the free energy difference between the liquid and the solid states at the temperature $T$. If the undercooling is small, one can ignore the difference in the specific heats of the liquid and the solid. The nucleation free energy is then proportional to the latent heat of fusion per atom $L$ [16],

$$\Delta G^{\text{nuc}} = L\frac{T - T_m}{T_m}. \tag{29}$$

When the undercooling is large, the expression (29) may be not precise enough. One can then consider the next term in the Taylor expansion [22] leading to

$$\Delta G^{\text{nuc}} = L\frac{T - T_m}{T_m} - \Delta C_p \frac{(T - T_m)^2}{2T_m}, \tag{30}$$

where $\Delta C_p = C_p^{\text{liq}}(T_m) - C_p^{\text{sol}}(T_m)$ is the difference in the heat capacities of the liquid and the solid phases.

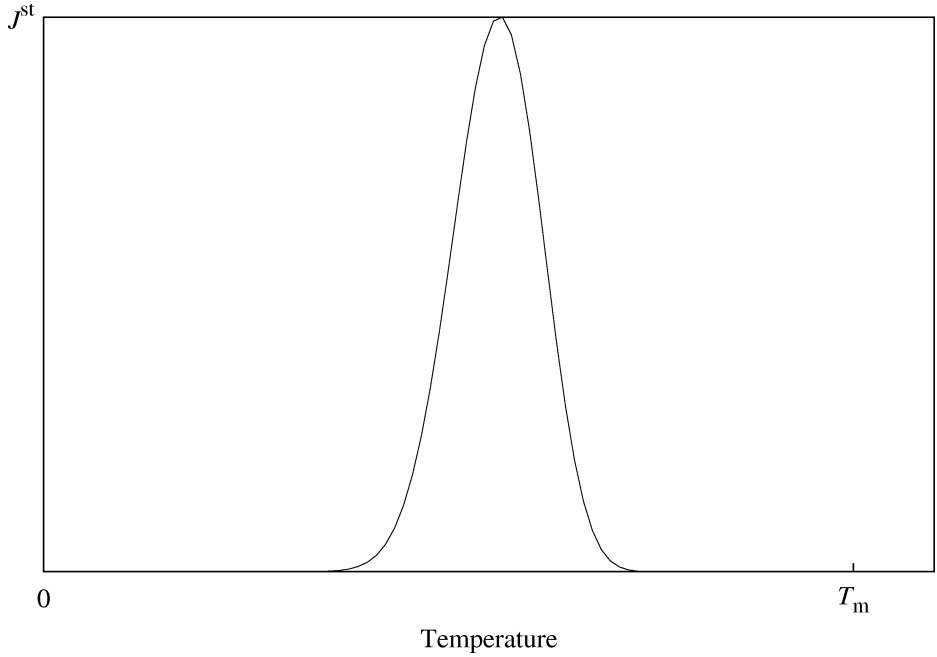

**Figure 7:** Variation with the quenching temperature of the steady-state nucleation rate in the case of solidification (Eq. (31)). Melting temperature, $T_m$.

Equation (29) to shows how the steady-state nucleation rate varies with the quenching temperature. Assuming that the interface free energy is constant and that the condensation rate $\beta^*$ simply obeys an Arrhenius law, one obtains that the nucleation rate is given by

$$J^{st} = \lambda \frac{(T-T_m)^2}{\sqrt{T}} \exp\left[-\left(\frac{A}{T} + \frac{B}{T(T-T_m)^2}\right)\right], \quad (31)$$

where $\lambda$, $A$ and $B$ are positive constants. The nucleation rate corresponding to this equation is sketched in Fig. 7. The main variations are given by the exponential. As a consequence, there is a temperature window in which the nucleation rate is substantial. For high temperatures close to the melting temperature $T_m$, the nucleation free energy is small and leads to a negligible nucleation rate. At low temperatures, the nucleation rate is also negligible because of the Arrhenian behavior of the kinetic factor and of the critical cluster concentration (Eq. (7)). The nucleation rate can be measured only at intermediate temperatures. Such conclusions on the nucleation rate are not specific to solidification but are encountered in any nucleation experiment.

### *Example 2: Precipitation in the Solid State*

In this example, it is necessary to take into account elastic effects. The free energy is thus divided between a chemical and an elastic contribution.

***Chemical contribution.*** For the binary mixture which free energy per atom $G(x)$ is sketched in Fig. 1, the homogeneous metastable phase of composition $x^0$ has a nucleation free energy given by

$$\Delta G^{nuc}(x^0) = (1-y^e)\left[\mu_A(y^e) - \mu_A(x^0)\right] + y^e\left[\mu_B(y^e) - \mu_B(x^0)\right]. \quad (32)$$

A and B atom chemical potentials are respectively defined as the first derivatives of the total free energy with respect to the number $N_A$ and $N_B$ of A and B atoms. This leads to the following expressions

$$\mu_A(x) = G(x) - xG'(x),$$
$$\mu_B(x) = G(x) + (1-x)G'(x), \quad (33)$$

which check the property $(N_A + N_B)G(x) = N_A\mu_A + N_B\mu_B$.

Incorporating these expressions in (32):

$$\Delta G^{nuc}(x^0) = G(y^e) - G(x^0) - (y^e - x^0)G'(x^0). \tag{34}$$

This shows that the nucleation free energy corresponds to the difference, calculated in the point of abscissa $y^e$, between the free energy and the tangent in $x^0$ as illustrated in Fig. 1.

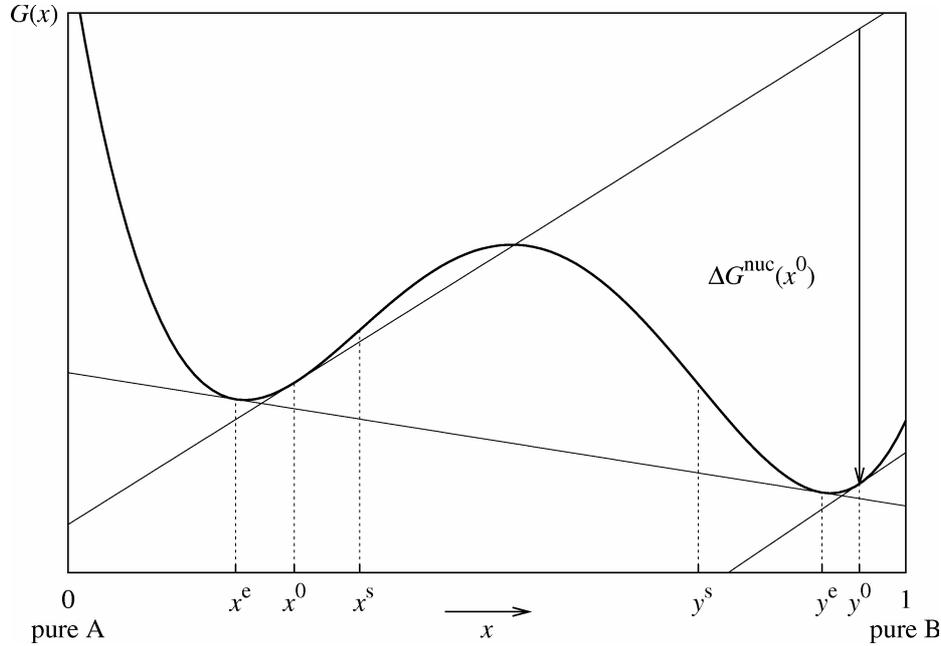

**Figure 8:** Parallel tangent construction leading to the maximal nucleation driving force for precipitates having the composition $y^0$.

It should be stressed, however, that this construction does not corresponds to the maximal nucleation driving force. If the stoichiometry of the precipitates is allowed to vary, the maximal nucleation driving force is obtained for a cluster composition $y^0$ corresponding to the point where the tangent to the free energy is parallel to the tangent in point of abscissa $x^0$ (Fig. 8). Such a deviation of the nucleating phase from its equilibrium may be important to consider. An example is the precipitation of carbonitride precipitates in steels [33]. In the following, we will consider that the free energy well defining the nucleating phase is deep enough so that the compositions $y^e$ and $y^0$ can be assumed identical. We will return to this question of the precipitate composition in the Kinetic Approach section, Non-stoichiometric Clusters, where a general framework to treat variations of the precipitate composition will be presented.

To go further, one needs to consider a precise function for the free energy. The regular solid solution is a convenient energetic model that is representative of a binary alloy. In this model, the free energy per atom is

$$G(x) = kT\left[x\ln(x) + (1-x)\ln(1-x)\right] + x(1-x)\omega, \tag{35}$$

where $\omega$ is the interaction parameter. When this parameter is positive, the alloy tends to unmix at low temperature and the corresponding phase diagram possesses a two-phase region. For temperatures lower than $\omega/2k$, the free energy has indeed two minima and its variation with the composition is similar to the one sketched in Fig. 1. The nucleation free energy of a solid solution quenched in a metastable state, as given by this thermodynamic model, is

$$\Delta G^{nuc}(x^0) = (1-y^e)kT\ln\left(\frac{1-x^e}{1-x^0}\right) + y^e kT\ln\left(\frac{x^e}{x^0}\right) + \omega(x^0 - x^e). \tag{36}$$

A useful approximation of this expression is the dilute limit corresponding to a small solubility limit, $x^{\text{e}} \ll 1$, and a small nominal concentration, $x^0 \ll 1$. In that case, one can keep only the major contribution in the nucleation free energy leading to

$$\Delta G^{\text{nuc}}\left(x^0\right) = y^{\text{e}} kT \ln\left(\frac{x^{\text{e}}}{x^0}\right). \tag{37}$$

This generally gives a good approximation of the nucleation free energy at low temperature for not too high supersaturations. It then allows predicting the main consequences of a variation of the solid solution nominal composition on the nucleation. This approximation of the nucleation free energy in the dilute limit can be easily generalized to a multi-component alloy.

Other thermodynamic approaches can be used to obtain expressions of the nucleation free energy. It is possible for instance to describe interactions between atoms with an Ising model. Chemical potentials entering in Eq. (4) can then be calculated with the help of current thermodynamic approximations, like mean-field approximations and low or high temperature expansions [34]. Using the simple Bragg-Williams mean field approximation one recovers indeed the expression (36) corresponding to the regular solid solution model. An example of this approach starting from an atomic model is given in references 17 and 35 for face-centered-cubic solid solution leading to the nucleation of a stoichiometric compound with the $L1_2$ structure like Al-Zr or Al-Sc alloys. On the other hand, it is possible to use experimental thermodynamic database, like the ones based on the Calphad approach [36, 37], to calculate the nucleation free energy.

*Elastic contribution.* Usually the precipitating phase has a different structure or molar volume from the parent phase. If the interface between both phases remains coherent, an elastic contribution needs to be taken into account in the formation free energy of the clusters (Eq. (3)). Like the "chemical" nucleation free energy, this elastic contribution varies linearly with the volume $V$ of the cluster. Its sign is always positive because there is an extra energy cost to maintain coherency at the interface. One can illustrate this elastic contribution by considering the case of a precipitating phase having a slightly different equilibrium volume from the parent phase as well as different elastic constants. For the sake of simplicity the assumption is that both phases have an isotropic elastic behavior characterized by their Lamé coefficients, $\lambda$ and $\mu$ for the parent phase, $\lambda'$ and $\mu'$ for the precipitating phase. If $a$ and $a(1+\delta)$ are the respective lattice parameters of the two phases, the elastic energy necessary to embed a spherical cluster of volume $V$ in an infinite elastic medium corresponding to the parent phase is

$$\Delta G^{\text{el}} = V \frac{6\mu(3\lambda' + 2\mu')}{3\lambda' + 2\mu' + 4\mu} \delta^2, \tag{38}$$

The model of the elastic inclusion and inhomogeneity developed by Eshelby [38, 39, 40] allows calculating the elastic energy in more complicated situations, when the inhomogeneity elastic behavior is anisotropic or when the inclusion stress-free strain is different from a simple pure dilatation. One can also deduce from this model the cluster shape minimizing its elastic self energy. Nevertheless, this model is tractable only when the inclusion is an ellipsoid. When the elastic contribution becomes important compared to the interface one, the shape of the critical cluster strongly deviates from an ellipsoid. One can then use a diffuse interface phase field model to determine the critical nucleus morphology and determine the associated nucleation activation energy [41, 42]. But in all cases, the extra energy cost arising from elasticity is positive and proportional to the inclusion volume. It thus reduces the absolute value of the nucleation driving force.

This inclusion model allows deriving the cluster self elastic energy. But the interaction of the cluster with the surrounding microstructure is ignored. In particular one does not consider the elastic interaction between different clusters. Such an interaction is long range and cannot always be neglected. It may lead to self-organized morphological patterns due to preferred nucleation sites around already existing clusters. In the case where the strain induced by the microstructure varies slowly compared to the size of the nucleating cluster, it has been shown that the interaction elastic energy depends linearly on the cluster volume and is independent on its shape [43]. This interaction energy, which sign is not fixed, depends on the position of the cluster. It can be considered in the cluster formation free energy (Eq. (3)) so as to model strain enhanced nucleation. Such a model is able to predict for instance variation of the nucleation driving force near an existing precipitate between the elastically soft and hard directions. A natural way to develop such a modeling is to use a phase field approach (See the Appendix to this article).

# Kinetic Approach

Predictions of the classical nucleation theory, *i.e.* the steady-state nucleation rate and the incubation time, are approximated solutions of kinetic equations describing the time evolution of the system. Instead of using results of the classical nucleation theory, one can integrate these kinetic equations numerically. This kinetic approach is known as the cluster dynamics. It rests on the description of the system undergoing phase separation as a gas of clusters which grow and decay by absorbing and emitting other clusters. In this section, the cluster gas thermodynamic formalism used by cluster dynamics is described first. Kinetic equations simulating the phase transformation are then presented. Finally, the link with classical nucleation theory is shown. It is generally assumed that the stoichiometry of the nucleating phase cannot vary. This is thus equivalent to considering the nucleation of clusters with a fixed composition which is known *a priori*. We will see at the end of this section how this strong assumption can be removed when one is interested in the nucleation of a multi-component phase with a varying composition.

### *Cluster Gas Thermodynamics*

The system is described as a gas of non-interacting clusters having a fixed stoichiometry corresponding to the one of the precipitating phase at equilibrium with the parent phase. Clusters are groups of atoms that are linked by a neighborhood relation. If one wants to model precipitation in an unmixing alloy for instance, one can consider that all solute atoms that are closer than a cutoff distance belong to the same cluster. No distinction is made between clusters belonging to the old or to the new phase. In this modeling approach, clusters are defined by a single parameter, their size or the number $n$ of atoms they contain. We call $G_n$ the free energy of a cluster containing $n$ atoms embedded in the solvent. $G_n$ is a free energy and not simply an energy because of the configurational entropy: for a given cluster size, there can be different configurations having different energies. One thus needs to consider the associated partition function. If $D_n^i$ is the number of configurations having the energy $H_n^i$ for a cluster of size $n$, the cluster free energy is then defined as

$$G_n = -kT \ln\left[\sum_i D_n^i \exp\left(-H_n^i/kT\right)\right]. \tag{39}$$

It is formally possible to divide this free energy in a volume and an interface contribution like in the capillary approximation, except that the interface free energy $\sigma_n$ may now depend on the cluster size. This free energy corresponds to an interface between the stoichiometric cluster and the pure solvent. Thus in three dimensions:

$$G_n = n g^e + n^{2/3}(36\pi\Omega^2)^{1/3}\sigma_n, \tag{40}$$

where $g^e$ is the free energy per atom of the bulk equilibrium precipitating phase, *i.e.* without any interface. This is, by definition, the sum of the chemical potentials $\mu_i^e$ for each constituent of the cluster modulated by its atomic fraction $y_i^e$,

$$g^e = \sum_i y_i^e \mu_i^e. \tag{41}$$

The interface free energy $\sigma_n$ entering in Eq. (40) is an average isotropic parameter and clusters, on average, are therefore assumed to be spherical. One important difference with the capillary approximation is that this interface free energy now depends on the size $n$ of the cluster. It is possible to compute the cluster free energy $G_n$ starting from an energetic model describing interactions between atoms. For small clusters, one can directly enumerate the different configurations $i$ accessible to a cluster of size $n$, and then directly build the free energy (Eq. (39)) [17, 32, 44]. As the degeneracy $D_n^i$ grows very rapidly with the size of the cluster, this approach is limited to small clusters. For larger clusters, one can sample thermodynamic averages with Monte Carlo simulations so as to compute the free energy difference between a cluster of size $n$ and one of size $n+1$ at a given temperature [44, 45]. These simulations have shown that, in 3 dimensions, the size dependence of the interface free energy is well described for large enough clusters by a generalized capillary approximation,

$$\sigma_n = \sigma\left(1 + cn^{-1/3} + dn^{-2/3} + en^{-2/3}\ln(n)\right), \tag{42}$$

where the temperature dependent coefficients $c$, $d$ and $e$ corresponds to the "line", the "point" and the "undulation" contributions to the interface free energy [45]. They take into account the interface curvature. The asymptotic limit of Eq. (42) corresponds to the constant interface free energy of the classic capillary approximation, which also depends on temperature.

Some other expressions have been proposed in the literature for the size dependence of the interface free energy. Gibbs [46] indeed obtained a differential equation of this size dependence. Integrating this expression, Tolman [47] obtained the following expression:

$$\sigma_n = \sigma\left[1 + \left(\frac{n_0}{n}\right)^{1/3}\right]^{-2}, \tag{43}$$

where $n_0$ is a parameter. One can see however that Eq. (42) and (43) are equivalent up to the order $o(n^{-1/3})$ when $n$ tends to infinity.

Consider an assembly composed of non interacting clusters. Thermodynamics is modeled in the cluster gas approximation of Frenkel [48]. If $N_n$ is the number of clusters containing $n$ atoms, the free energy of the system is given by

$$G = G_0 + \sum_{n=1}^{\infty} N_n G_n - kT \ln(W), \tag{44}$$

where $G_0$ is the free energy in the absence of clusters and $W$ is the number of different configurations accessible to the cluster assembly. Assuming that each cluster, whatever its size, lies only on one site and neglecting around each cluster all excluded sites which cannot be occupied by any other cluster, this number is simply given by

$$W = \frac{N_0!}{\left(N_0 - \sum_{n=1}^{\infty} N_n\right)! \prod_{n=1}^{\infty} N_n!}, \tag{45}$$

where $N_0$ is the number of sites accessible to the cluster. Application of the Stirling formula leads to the following estimation for the free energy

$$G = G_0 + \sum_{n=1}^{\infty} N_n G_n + kT \sum_{n=1}^{\infty} N_n \ln(N_n) + kT\left(N_0 - \sum_{n=1}^{\infty} N_n\right)\ln\left(N_0 - \sum_{n=1}^{\infty} N_n\right) - N_0 \ln(N_0). \tag{46}$$

We can deduce from this free energy the equilibrium cluster size distribution. This distribution is obtained by minimizing (46) under the constraint that the total number of atoms included in the clusters is fixed. We therefore introduce a Lagrange multiplier $\mu$ and define the grand canonical free energy

$$G - \mu \sum_{n=1}^{\infty} n N_n. \tag{47}$$

The minimization of this grand canonical free energy with respect to the variables $N_n$ leads to the equilibrium cluster size distribution which should check the equation

$$\frac{N_n^{eq}}{N_0 - \sum_{n=1}^{\infty} N_n^{eq}} = \exp\left(-\frac{G_n - n\mu}{kT}\right). \tag{48}$$

The assumption of non interacting clusters used to derive this equation is only valid in the dilute limit. It is therefore reasonable to neglect in Eq. (48) the sum appearing in the right hand side denominator compared to the number of accessible sites $N_0$. At equilibrium, the atomic fraction of clusters containing $n$ atoms is then

$$C_n^{\text{eq}} = \frac{N_n^{\text{eq}}}{N_s} = C_0 \exp\left(-\frac{G_n - n\mu}{kT}\right), \quad (49)$$

where $C_0 = N_0 / N_s$ and $N_s$ is the total number of sites. For homogeneous nucleation, all sites can act as nucleation centers: $N_0 = N_s$ and $C_0 = 1$. Sometimes, Eq. (49) is written in its equivalent form:

$$C_n^{\text{eq}} = C_0 \left(\frac{C_1^{\text{eq}}}{C_0}\right)^n \exp\left(-\frac{G_n - nG_1}{kT}\right). \quad (50)$$

The quantities $G_n - n\mu$ and $G_n - nG_1$ should nevertheless not be confused: the first one is the cluster formation free energy in a cluster gas characterized by the parameter $\mu$, whereas the last one is the energy difference between the cluster and the equivalent number of monomers. In the following, we will rather use Eq. (49) as it allows to make a direct link with the capillary approximation used by the classical nucleation theory.

It is interesting to understand the physical meaning of the Lagrange multiplier $\mu$ appearing in (49). At equilibrium, the grand canonical free energy (47) is at a minimum. We must then have for all sizes $n$:

$$\mu = \frac{1}{n} \frac{\partial G}{\partial N_n}. \quad (51)$$

To calculate this derivative, we introduce the total number of atoms of type $i$

$$M_i = y_i^{\text{e}} \sum_{n=1}^{\infty} n N_n. \quad (52)$$

Eq. (51) is equivalent to

$$\begin{aligned} \mu &= \frac{1}{n} \sum_i \frac{\partial G}{\partial M_i} \frac{\partial M_i}{\partial N_n}, \\ &= \sum_i \mu_i^0 y_i^{\text{e}} \end{aligned} \quad (53)$$

where we have used the definition of the chemical potential – first derivative of the total free energy with respect to the number of atoms. One therefore sees that the Lagrange multiplier is nothing else than the chemical potentials of the different atomic species modulated by their atomic fraction. The fact that only one Lagrange multiplier is needed and not one for each constituent is a consequence of our initial assumption that the clusters have a fixed composition corresponding to the equilibrium one $y_i^{\text{e}}$. Using the expression (40) of the cluster free energy and the definition (41) of the volume contribution, one recovers the equilibrium cluster size distribution given by the capillary approximation:

$$C_n^{\text{eq}} = C_0 \exp\left(-\frac{\Delta G_n}{kT}\right), \quad (54)$$

with

$$\Delta G_n = n \Delta G^{\text{nuc}} + n^{2/3} \left(36\pi \Omega_1^2\right)^{1/3} \sigma_n. \quad (55)$$

The nucleation free energy has the same expression as the one used in classical nucleation theory (Eq. (4)), but now the interface free energy depends on the cluster size.

We would like to stress that the cluster gas approximation is a thermodynamic model by itself: thermodynamic quantities like chemical potentials are results and not input parameters of the model [49]. This will have important consequences for the kinetic approach of nucleation developed in the next section: in contrast with classical nucleation theory, one will not need to calculate the nucleation driving force to input it in the modeling.

One can use this cluster gas thermodynamic model to calculate the composition of the parent phase at the coexistence point between the parent and the nucleating phase, *i.e.* the solubility limit.

This coexistence point is defined by the equality of the chemical potentials $\mu_i^0$ and $\mu_i^e$. The nucleation free energy is thus null and only the interface contributes to the cluster formation free energy (55). At the coexistence point, the composition of the parent phase is then

$$x_i^e = y_i^e \sum_{n=1}^{\infty} n \exp\left(-\frac{n^{2/3}(36\pi\Omega_1)^2 \sigma_n}{kT}\right). \tag{56}$$

We see that the interface free energy fixes the solubility limit in the parent phase. This interface free energy is actually the key parameter of the nucleation kinetic approach. Even if its dependence with the cluster size is small, it is generally important to take it into account as all thermodynamic quantities derive from it and it enters in exponential terms like in Eq. (56).

### *Cluster Dynamics*

For the sake of simplicity, in the following subsections homogeneous nucleation is considered. All monomers can be assumed equivalent: one does not need to distinguish between monomers lying on nucleation sites and free monomers.

#### *Master Equation*

Kinetics is described thanks to a master equation which gives the time evolution of the cluster size distribution. In a lot of cases, one can assume that only monomers migrate. We will therefore first consider this assumption and consider latter the case where all clusters are mobile. When only monomers can migrate, the probability to observe a cluster containing $n$ atoms obeys the differential equations:

$$\frac{\partial C_n}{\partial t} = J_{n-1 \to n} - J_{n \to n+1} \quad \forall n \geq 2,$$
$$\frac{\partial C_1}{\partial t} = -2J_{1 \to 2} - \sum_{n \geq 2} J_{n \to n+1}, \tag{57}$$

where $J_{n \to n+1}$ is the cluster flux from the class of size $n$ to the class $n+1$. This flux can be written

$$J_{n \to n+1} = \beta_n C_n - \alpha_{n+1} C_{n+1}, \tag{58}$$

where $\beta_n$ is the probability per unit time for one monomer to impinge on a cluster of size $n$ and $\alpha_n$ for one monomer to leave a cluster of size $n$.

#### *Condensation Rate*

Expression of the condensation rate $\beta_n$ can be obtained from physical considerations. This condensation rate has to be proportional to the monomer concentration and one can generally write

$$\beta_n = b_n C_1, \tag{59}$$

where $b_n$ is an intrinsic property of the cluster of size $n$. In the ballistic regime, this factor is proportional to the surface of the cluster and to the jump frequency $\Gamma_1$ of the monomer to impinge on the cluster. In the diffusion regime, this factor is proportional to the cluster radius and to the monomer diffusion coefficient $D_1$. A general expression of the condensation rate, covering the ballistic and the diffusion regime, has been proposed by Waite [25] who obtained

$$\beta_n = 4\pi \frac{R_n^2}{R_n + \kappa} \frac{D_1}{\Omega_1} C_1, \tag{60}$$

where $\Omega_1$ is the monomer volume and $R_n$ is the cluster capture radius. One can assume that this radius is close to the one corresponding to the more compact cluster shape, *i.e.* a sphere, leading to

$$R_n = \left(\frac{3n\Omega_1}{4\pi}\right)^{1/3}. \tag{61}$$

The distance $\kappa$ is given by the relation

$$\kappa = \frac{D_1}{\lambda_1 \Gamma_1}, \qquad (62)$$

where $\lambda_1$ is the distance corresponding to the monomer last jump to become attached to the cluster. If $R_n \ll \kappa$, one recovers the expression of the condensation rate in the ballistic regime, and in the diffusive regime if $R_n \gg \kappa$. Eq. (60) therefore shows that condensation on small cluster is generally controlled by ballistic reactions, and condensation on big clusters by diffusion.

The expressions used by classical nucleation theory for the condensation rate (Eq. (12) and (13)) are similar to the ballistic and diffusion limits of Eq. (62). Nevertheless, a difference appears as the condensation rate of the classical nucleation theory is proportional to the solute concentration and not to the monomer concentration like in Eq. (62). It makes thus use of the total solute diffusion coefficient or jump frequency and not of the monomer diffusion coefficient or jump frequency. For a dilute system, one can consider that all the solute is contained in monomers. The condensation rates used by both approaches are then equivalent. But the difference may be important for more concentrated systems. This point has been thoroughly discussed by Martin [49] who showed the equivalence in the dilute limit.

### *Evaporation Rate*

By contrast with the condensation rate, the evaporation rate $\alpha_n$ cannot be generally obtained directly. It has to be deduced from $\beta_n$ using the equilibrium cluster size distribution (49). The evaporation rate is obtained assuming that it is an intrinsic property of the cluster and does not depend on the embedding system. We therefore assume that the cluster has enough time to explore all its configurations between the arrival and the departure of a monomer. This assumption is coherent with the fact that we only describe the clusters through their sizes. $\alpha_n$ should thus not depend on the saturation of the embedding system. It could be obtained in particular by considering any undersaturated system. Such a system is stable and there should be then no energy dissipation. This involves all fluxes $J_{n \to n+1}$ equaling zero. Using Eq. (58), one obtains

$$\alpha_{n+1} = \bar{\alpha}_{n+1}(\mu) = \bar{\beta}_n(\mu) \frac{\bar{C}_n(\mu)}{\bar{C}_{n+1}(\mu)}, \qquad (63)$$

where overlined quantities are evaluated in the system at equilibrium characterized by its effective chemical potential $\mu$. In particular, the cluster size distribution is the equilibrium one given by Eq. (49). Using the expression (59) of the condensation rate, this finally leads to the following expression of the evaporation rate:

$$\alpha_{n+1} = b_n C_0 \exp\left[ (G_{n+1} - G_n - G_1)/kT \right]. \qquad (64)$$

As the condensation rate varies linearly with the monomer concentration, the contribution of the effective chemical potential cancels out in the expression (63) of $\alpha_n$. We recover our starting assumption: the evaporation rate only depends on the cluster free energy and not on the overall state of the cluster gas characterized by the effective chemical potential $\mu$. Using the generalized capillary approximation (55), one can show that the evaporation rate actually only depends on the cluster interface free energy:

$$\alpha_{n+1} = b_n C_0 \exp\left\{ \left( 36\pi\Omega_1^2 \right)^{1/3} \left[ (n+1)^{2/3} \sigma_{n+1} - n^{2/3} \sigma_n - \sigma_1 \right]/kT \right\}. \qquad (65)$$

The evaporation rate is then independent of the nucleation free energy $\Delta G^{\text{nuc}}$ which does not appear in any parameter. The nucleation free energy is implicit in cluster dynamics: there is no need to know it but, if needed, one can calculate it from the cluster gas thermodynamic. This is to contrast with classical nucleation theory where the nucleation free energy is an input parameter. On the other hand, cluster dynamics is very sensitive on the interface free energy as it appears in an exponential in the expression (65) of the evaporation rate. It is very important to have a correct evaluation of this interface free energy, especially of its variations with the cluster size, at least for small sizes.

In our approach, the evaporation rate is derived assuming that it is an intrinsic property of the cluster. Sometimes, one derives this parameter assuming instead that a hypothetical constrained equilibrium exists for the clusters in the supersaturated system: the equilibrium cluster size distribution (49) is taken to hold although the system is supersaturated and cannot be at equilibrium. The evaporation rate is then obtained by imposing detailed balance for Eq. (58) with respect to this constrained equilibrium. Comparison with atomic simulation of the magnetization reversal of an Ising model [20] has shown that this constraint equilibrium assumption is good. The same conclusion was reached for subcritical clusters in the case of precipitation in the solid state [50]. Katz and Wiedersich [51] pointed out that this "constrained equilibrium" assumption generally leads to the same expression of the evaporation rate as our "intrinsic property" assumption. In particular, this is true when the condensation rate varies linearly with the monomer concentration as this is the case here (Eq. (59)).

When the growth and decay of clusters is controlled by a reaction at the interface (ballistic regime), it is also possible to compute directly the condensation and evaporation rates [44]. An atomistic model is used to describe the physical process at the atomic scale and the corresponding rates are obtained by thermal averaging through Monte Carlo sampling. Detailed balance is now imposed at the atomic scale. This ensures that the detailed balance at the cluster scale as given by Eq. (63) is also checked. A huge computational effort is required, but this could be optimized by calculating at the same time the cluster interface free energies and their condensation and evaporation rates.

*Numerical Scheme*

The evolution of the cluster size distribution is obtained by integrating the set of equations (57). A direct approach can become cumbersome as the number of differential equations varies linearly with the size of the largest cluster. The maximum size of the cluster that can be considered is therefore limited by the number of differential equations that can be integrated. This problem can be circumvented by noticing that a detailed description is important only for small cluster sizes where quantities vary rapidly. For large sizes, variations are smoother and an approximated description can be used. The easiest approach to do so is to consider that the size $n$ is now a continuous variable. One can then develop Eqs. (57) and (58) to the second order about $n$ and the system evolution is described by the Fokker-Planck equation [52]:

$$\frac{\partial C_n}{\partial t} = -\frac{\partial}{\partial n}\left[(\beta_n - \alpha_n)C_n\right] + \frac{1}{2}\frac{\partial^2}{\partial n^2}\left[(\beta_n + \alpha_n)C_n\right]. \tag{66}$$

This continuous equation can be solved numerically by discretizing the continuous variable $n$. The best way to handle large cluster sizes is to use a varying increment greater than 1 and increasing with the cluster size. A convenient solution is an increment growing at a constant rate $\lambda$. The variable $n$ is then discretized according to

$$\begin{aligned} n_j &= j, \quad \forall j \leq n_d \\ n_j &= n_d + \frac{1-\lambda^{j-n_d}}{1-\lambda}, \quad \forall j \geq n_d, \end{aligned} \tag{67}$$

where $n_d$ is the number of classes for which the discrete equation (57) is used. Above this size, one integrates instead the discretized version of Eq. (66) which is:

$$\begin{aligned} \frac{\partial C_{n_j}}{\partial t} &= \frac{1}{n_{j+1}-n_{j-1}}\left[(\beta_{n_j}-\alpha_{n_j}) + \frac{\beta_{n_j}+\alpha_{n_j}}{n_j-n_{j-1}} - \frac{\partial}{\partial n}(\beta_{n_j}+\alpha_{n_j})\right]C_{n_{j-1}} \\ &+ \left[-\frac{\beta_{n_j}+\alpha_{n_j}}{(n_{j+1}-n_j)(n_j-n_{j-1})} - \frac{\partial}{\partial n}(\beta_{n_j}-\alpha_{n_j}) - \frac{1}{2}\frac{\partial^2}{\partial n^2}(\beta_{n_j}+\alpha_{n_j})\right]C_{n_j} \\ &+ \frac{1}{n_{j+1}-n_{j-1}}\left[-(\beta_{n_j}-\alpha_{n_j}) + \frac{\beta_{n_j}+\alpha_{n_j}}{n_{j+1}-n_j} + \frac{\partial}{\partial n}(\beta_{n_j}+\alpha_{n_j})\right]C_{n_{j+1}}, \end{aligned} \tag{68}$$

The evolution of the monomer concentration is approximated by

$$\frac{\partial C_1}{\partial t} = -2\beta_1 C_1 + \alpha_2 C_2 + \sum_{j\geq 2}^{n_d}\left(\alpha_{n_j} - \beta_{n_j}\right)C_{n_j} + \sum_{j\geq n_d}\left(\alpha_{n_j} - \beta_{n_j}\right)\frac{n_{j+1} - n_{j-1}}{2}C_{n_j} . \tag{69}$$

This numerical scheme is simple and allows reaching large cluster sizes with a reasonable number of differential equations. Typically, it is possible to simulate clusters containing more than 4 million atoms by using 100 discrete classes and 400 continuous classes with a growing increment rate $\lambda = 1.03$. One should nevertheless mention that this numerical scheme does not strictly conserve the matter. By using reasonable values for the discretization parameters $\lambda$ and $n_d$ the losses are generally insignificant, but, in any cases, they must be checked afterwards to see if they are acceptable. One has to verify also that the concentration of the largest size has not evolved at the end of the simulation.

Another numerical approach has been proposed by Kiritani [53] to solve the set of differential equations (57) while allowing reaching large cluster size. His grouping method consists in replacing a group of master equations by only one equation representing the class. It assumes that the number of clusters of each size in a group is the same and that the condensation and evaporation rates for clusters in a group do not vary. It has been shown unfortunately that the result can be very bad if the grouping is not carried out properly [54]. Furthermore, as the previous scheme, it does not conserve strictly the matter even with an optimized grouping. Golubov *et al.* [55] proposed a new grouping method that can conserve the matter. For this purpose, the first and the second moments of each group are considered and two equations for each class are obtained. The first moment equation controls the time evolution of the cluster size distribution and the second moment equation ensures the matter conservation. Such a numerical scheme therefore requires twice more equations than the one proposed above.

### The Link with Classical Nucleation Theory

Main results of classical nucleation theory have been actually derived from cluster dynamics, *i.e.* from the master equation (57) describing the time evolution of the cluster population. This derivation is interesting as it allows a better understanding of the assumptions behind the classical nucleation theory. Moreover, it gives insights how this theory can be further developed to broaden the range where it applies. In the following sub-section, we fist compare the definition of the critical size in cluster dynamics with the classical ones, and then see how the steady-state nucleation rate and the corresponding cluster size distribution can be derived from the master equation. The derivation of the incubation time is not given here but can be found in Refs. 26 and 27 for instance.

#### *Critical Size*

Sub-critical clusters are unstable: they have a higher probability to decay than to grow. On the contrary, super-critical clusters are stable and have a higher probability to grow than to decay. The critical size $n*$ is then defined as the size for which the condensation rate equals the evaporation rate:

$$\beta_{n*} = \alpha_{n*} . \tag{70}$$

This definition is actually different from the one used by the classical nucleation theory where the critical size is the size for which the cluster formation free energy is maximum. One can show that these two definitions are consistent and lead to the same expression in the limit of large cluster sizes. To do so, rewrite Eq. (70) using the expressions (59) of the condensation rate and (64) of the evaporation rate:

$$b_{n*}C_1 = b_{n*-1}C_0 \exp\left(\frac{G_{n*} - G_{n*-1} - G_1}{kT}\right) . \tag{71}$$

Then assume that monomers are at local equilibrium: their concentration $C_1$ obeys the equilibrium cluster size distribution (49). One can thus eliminate in Eq. (71) the monomer free energy $G_1$:

$$b_{n*}\exp\left(\frac{\mu}{kT}\right) = b_{n*-1}\exp\left(\frac{G_{n*} - G_{n*-1}}{kT}\right) . \tag{72}$$

Using the definition $\Delta G_n = G_n - n\mu$ of the cluster formation free energy, Eq. (72) can be rewritten

$$\frac{b_{n*-1}}{b_{n*}} \exp\left(\frac{\Delta G_{n*} - \Delta G_{n*-1}}{kT}\right) = 1. \tag{73}$$

Using Eq. (55) to express the cluster formation free energy, one finally obtain that the critical size verifies

$$\frac{b_{n*-1}}{b_{n*}} \exp\left\{\frac{\Delta G^{\text{nuc}} + \left(36\pi\Omega_1^2\right)^{1/3}\left[n*^{2/3}\sigma_{n*} - (n*-1)^{2/3}\sigma_{n*-1}\right]}{kT}\right\} = 1. \tag{74}$$

To go further, one needs to take the limit corresponding to large cluster sizes. One can then neglect the size dependence of the condensation rate prefactor, $b_{n*-1} \sim b_{n*}$, and of the cluster interface free energy, $\sigma_{n*-1} \sim \sigma_{n*} \sim \sigma$. At the critical size, one should therefore check

$$\Delta G^{\text{nuc}} + \left(36\pi\Omega_1^2\right)^{1/3}\left[n*^{2/3} - (n*-1)^{2/3}\right]\sigma = 0. \tag{75}$$

A limited expansion of Eq. (75) for large sizes leads to the result:

$$n* = \left[-\frac{2}{3}\frac{\left(36\pi\Omega_1^2\right)^{1/3}\sigma}{\Delta G^{\text{nuc}}}\right]^3. \tag{76}$$

One therefore recovers Eq. (5) of the critical size with a geometric factor $A$ corresponding to spherical clusters. The critical size considered by classical nucleation theory corresponds to the one of cluster dynamics in the limit of large cluster sizes. However, when the critical size is small, both definitions may differ. This coherence of both definitions at large size and this deviation at small sizes has been observed in atomic simulations [20].

### *The Steady-State Nucleation Rate*

One can calculate the steady-state nucleation rate $J^{\text{st}}$ corresponding to the master equation (57). To do so, one needs to make two assumptions:
- There is a small size below which clusters have their equilibrium concentration given by Eq. (49). Clusters smaller than the critical size appear and disappear spontaneously through thermal fluctuations and their concentrations stay roughly at equilibrium. The smaller the cluster, the better this assumption. The most convenient choice is therefore to impose thermal equilibrium for monomers:

$$C_1(t) = C_1^{\text{eq}} = C_0 \exp\left(-\frac{G_1 - \mu(t)}{kT}\right) \tag{77}$$

- There is a maximum cluster size $N$ above which the cluster concentration remains null: $C_N(t) = 0$. This assumption cannot be checked for a true steady-state without invoking a demon which removes clusters that appear at the size $N$ and dissolves them into monomers. Nevertheless, one can always define at a given time a size large enough so that the cluster distribution did not propagate to this size.

By definition, the steady-state nucleation rate can be calculated at any cluster size $n$. At the steady-state, all cluster concentrations remain constant. As a consequence

$$\frac{\partial J_{n \to n+1}}{\partial n} = 0, \tag{78}$$

and the steady-state nucleation rate can be calculated at any given cluster size. Using the expression (58) of the cluster flux with Eqs. (59) and (64) for the condensation and evaporation rates, one obtains

$$\begin{aligned} J^{\text{st}} &= b_n\left\{C_1 C_n - C_{n+1} C_0 \exp\left[\frac{G_{n+1} - G_n - G_1}{kT}\right]\right\}, \quad \forall n, \\ &= b_n C_1 \exp\left[-\frac{G_n - n\mu}{kT}\right]\left\{C_n \exp\left[\frac{G_n - n\mu}{kT}\right] - C_{n+1}\exp\left[\frac{G_{n+1} - (n+1)\mu}{kT}\right]\right\}. \end{aligned} \tag{79}$$

In this equation, we have used the fact that monomers are at equilibrium (Eq. (77)) to go from the first to the second line. After rearranging the terms between the left and right hand sides, we sum between a minimal and a maximal size

$$\sum_{n=n_1}^{n_2} \frac{J^{st}}{b_n C_1 C_0 \exp\left[-\dfrac{G_n - n\mu}{kT}\right]} = \frac{C_{n_1}}{C_0 \exp\left[-\dfrac{G_{n_1} - n_1\mu}{kT}\right]} - \frac{C_{n_2+1}}{C_0 \exp\left[-\dfrac{G_{n_2+1} - (n_2+1)\mu}{kT}\right]}. \tag{80}$$

We choose $n_1 = 1$ so that the first term in the right hand side is equal to 1. With $n_2 = N - 1$, the second term is null: we assumed $C_N = 0$ and the exponential is tending to $\infty$ for high enough $N$. We therefore arrive to the result

$$J^{st} = C_1 C_0 \frac{1}{\sum_{n=1}^{N-1} \dfrac{1}{b_n} \exp\left[\dfrac{G_n - n\mu}{kT}\right]}. \tag{81}$$

This gives an exact expression of the steady-state nucleation rate under both above assumptions.

The sum appearing in Eq. (81) can be easily evaluated. To do so, one makes a continuous approximation so as to transform the sum into an integral. One can further notice that the cluster formation free energy, $\Delta G_n = G_n - n\mu$, presents a maximum at the critical size $n^*$. As a consequence, the main contribution to the integral arises from sizes around the critical size and can be evaluated by a Taylor expansion around $n^*$. Finally, neglecting the variations of $b_n$ in front of the exponential, we obtain

$$J^{st} = C_1 C_0 b_{n^*} \frac{1}{\int_1^{N-1} \exp\left[\left(\Delta G_{n^*} + \dfrac{1}{2} \left.\dfrac{\partial^2 \Delta G_n}{\partial n^2}\right|_{n=n^*} (n - n^*)^2\right) \Big/ kT\right] dn}. \tag{82}$$

Changing the integration limits in $-\infty$ and $+\infty$, one recovers the result of classical nucleation theory

$$J^{st} = \beta^* Z C_0 \exp\left(-\frac{\Delta G^*}{kT}\right), \tag{83}$$

where $\beta^* = C_1 b_{n^*}$, $\Delta G^* = \Delta G_{n^*}$ and the Zeldovich factor is given by Eq. (9).

### *The Steady-State Cluster Size Distribution*

Once the steady-state nucleation rate is known, one can easily obtain the corresponding cluster size distribution. One uses again Eq. (80) with the limits $n_2 = N - 1$, so that the last term in the right hand side is still null, and with $n_1 = n$, the size at which we want to calculate the cluster concentration. This leads to the result

$$C_n^{st} = C_0 \exp\left[-\frac{G_n - n\mu}{kT}\right] \sum_{j=n}^{N} \frac{J^{st}}{b_n C_1 C_0 \exp\left[-\dfrac{G_j - j\mu}{kT}\right]}. \tag{84}$$

Like for the steady-state nucleation rate, the sum can be evaluated by making a continuous approximation, developing the cluster formation free energy around the critical size and considering the limit $N \to \infty$. One obtains

$$\begin{aligned}C_n^{st} &= C_0 \exp\left[-\frac{G_n - n\mu}{kT}\right] \frac{J^{st}}{C_1 C_0 b_{n^*}} \int_n^\infty \exp\left[\left(\Delta G_{n^*} + \frac{1}{2} \frac{\partial^2 \Delta G_j}{\partial j^2} (j - n^*)^2\right) \Big/ kT\right] dj \\ &= \frac{1}{2} \text{erfc}\left[\sqrt{\pi} Z (n - n^*)\right] C_0 \exp\left[-\frac{G_n - n\mu}{kT}\right].\end{aligned} \tag{85}$$

The stationary distribution therefore corresponds to the equilibrium one reduced by a factor varying from 0 for large sizes to 1 for the small sizes. Well below the critical size, *i.e.* for $n \leq n^* - \Delta n/2$ with $\Delta n$ given by Eq. (10), this factor only slightly differs from 1 and one recovers that the stationary distribution corresponds to the equilibrium one. At the critical size $n^*$, this factor is exactly one half and in the vicinity of $n^*$ one can approximate the stationary distribution with Eq. (8).

### *Discussion*

This derivation of quantities predicted by classical nucleation theory from cluster dynamics formalism enlightens the approximations made by this theory. It assumes that the supersaturation is not too high so that the critical size is large enough. This allows one to consider the size as a continuous instead of a discrete variable and to make a finite expansion of key parameters around the critical size. Classical nucleation theory may therefore appear as more restricted than the kinetic approach based on the master equation (57), but the situation is not so simple.

One severe restriction of the cluster dynamics is the thermodynamic model on which it relies. It is based on the cluster gas model of Frenkel [48], which is valid for a dilute system. Strictly speaking, cluster dynamics should only be used in the dilute case. If one wants to study more concentrated systems, one needs to extend the cluster gas model. Such an extension has been performed by Lépinoux [56] and is presented in the section, Configurational Frustrations between Clusters. On the other hand, classical nucleation theory does not rely on the cluster gas thermodynamic model. It makes use instead of the nucleation driving force which may be calculated with any thermodynamic model, in particular one better suited to concentrated systems. This is therefore not a problem to use the classical nucleation theory to study concentrated systems as long as one correctly calculated the nucleation driving force.

Both formalisms also differ in the way they describe the parent and the nucleating phases. In the classical theory, one differentiates both phases and nucleation is described through hetero-phase fluctuations corresponding to precipitates embedded in the parent phase. Such a differentiation does not appear in the kinetic approach where one only deals with one system which is described as a gas of clusters having a fixed stoichiometry and embedded in a pure solvent. This description difference may become relevant when modelling concentrated systems as the values of the input parameters may then differ according to the chosen, thermodynamic or kinetic, approach. This is the case for instance of the interface free energy. In cluster dynamics, this corresponds to the energy cost of an interface between a cluster with a fixed stoichiometry and the pure solvent, whereas in classical nucleation theory one should rather consider that the precipitate and the parent phase are not pure and that solubility exists in both phases. The concentration appearing in the expression of the condensation rate may also differ between both approaches as already quoted in the previous section on Condensation Rate. This is either the monomer concentration (cluster dynamics) or the total solute concentration (classical nucleation theory).

All these subtle differences between cluster dynamics and classical nucleation theory have been discussed by Martin [49] in the case of precipitation in the solid state. He showed that both approaches were consistent and lead to the same expressions in the dilute limit.

## Extensions of Cluster Dynamics

The master equation (57) can be modified so as to describe nucleation under less restricted conditions as the ones of the previous subsections, and then to build extensions of the cluster dynamics. In particular, the assumptions that only monomers can react and that clusters have a fixed stoichiometry corresponding to the equilibrium nucleating phase can be removed.

### *Mobile Clusters*

Until now we have assumed that only monomers are mobile. This assumption is not always valid. There is no reason to think for instance that all clusters except monomers are immobile in solidification. Diffusion of small clusters can also happen in solid-solid phase transformations. An interesting example is copper precipitation in iron where atomic simulations have revealed that clusters containing up to several tens of copper atoms can be much more mobile than individual copper atoms [57]. The master equation should therefore be modified so as to account for reactions involving clusters larger than monomers. Such a generalization of the cluster dynamics formalism has been performed by Binder and Stauffer [4, 5, 6].

The probability to observe a cluster containing $n$ atoms now obeys the generalized master equation

$$\frac{\partial C_n}{\partial t} = \frac{1}{2}\sum_{n'=1}^{n-1} J(n-n',n' \to n) - \sum_{n'=1}^{\infty} J(n,n' \to n+n') - J(n,n \to 2n), \qquad (86)$$

where $J(n,n' \to n+n')$ is the cluster flux corresponding to the reaction between the classes $n$ and $n'$ to the class $n+n'$. The factor $1/2$ appearing in Eq. (86) accounts for overcounting the pairs $\{n-n',n'\}$ in the summation. In this equation, one should not forget that the reactions $n+n \rightleftarrows 2n$ involve two clusters of size $n$. When reactions are limited to reactions involving monomers, $n'$ can only take the values $1$ and $n-1$ in the first sum, and $1$ in the second sum: one recovers the classical master equation (57) of cluster dynamics.

The cluster flux is the difference between the condensation of two clusters of sizes $n$ and $n'$ and the splitting of a cluster of size $n+n'$ in two clusters of sizes $n$ and $n'$,

$$J(n,n' \to n+n') = b(n,n' \to n+n')C_n C_{n'} - \alpha(n+n' \to n,n')C_{n+n'}. \qquad (87)$$

One should then get an expression for the absorption coefficient $b(n,n')$. In the case the reaction is limited by the cluster diffusion, this coefficient is given by [25]

$$b(n,n' \to n+n') = 4\pi R_{n,n'} \frac{D_n + D_{n'}}{\Omega_1}, \quad \forall n \neq n',$$

$$b(n,n \to 2n) = 4\pi R_{n,n} \frac{D_n}{\Omega_1}, \qquad (88)$$

where $R_{n,n'}$ is a capture radius and can be approximated by the sum of the two reacting clusters radii. In the expression (13) used in the classical nucleation theory, we identified this capture radius with the radius of the critical cluster and we neglected the monomer radius.

If the diffusion coefficients of the $n$-mers are not known, one can use an approximation proposed by Binder *et al.* [58, 59]. They simply consider that cluster diffusion is due to jumps of atoms located at the interface. When an atom jumps over a distance $r_s$ with a frequency $\Gamma_s$, the center of gravity of the cluster jumps over $r_s/n$. Since the number of possible jumps at the interface increases with its area as $n^{2/3}$, $D_n$ depends on $n$ as

$$D_n = \Gamma_s \left(\frac{r_s}{n}\right)^2 n^{2/3} = D_1 n^{-4/3}. \qquad (89)$$

In the case of precipitation in the solid state, one should not forget that substitutional atoms diffuse through exchange with vacancies and that a vacancy enrichment at the cluster interface is possible. In such a case, Eq. (89) has to be corrected with a prefactor so as to consider the vacancy concentration at the interface [57]. This vacancy segregation is the reason why clusters containing several copper atoms are more mobile than monomers in iron [57].

The evaporation rate is still obtained by assuming that it is an intrinsic property of the cluster (or imposing a constrained equilibrium), thus leading to

$$\alpha(n+n' \to n,n') = b(n,n' \to n+n')C_0 \exp\left(\frac{G_{n+n'} - G_n - G_{n'}}{kT}\right). \qquad (90)$$

All parameters are thus determined and the master equation (86) can be numerically integrated.

Binder and Stauffer [4, 6] also extended classical nucleation theory so as to get expressions of the steady-state nucleation rate and of the incubation time taking into account the mobility of all clusters. They started from the master equation (86) and imposed the detailed balance corresponding to Eq. (90). They obtained expressions similar to the classical ones, Eq. (11) for the steady-state nucleation rate and Eq. (14) for the incubation time, except that now the growing rate $\beta^*$ of the critical cluster incorporates contributions of all clusters. This growing rate is given by

$$\beta^* = \sum_{n=1}^{n_c} b(n^*, n \to n+n^*) n^2 C_0 \exp\left(-\frac{G_n - n\mu}{kT}\right), \tag{91}$$

where $n_c$ is a cut-off size corresponding to the correlation length of thermal fluctuations. It seems reasonable to identify this cut-off size with the critical size $n^*$. When only reactions involving monomers can occur, the sum in Eq. (91) is limited to the term $n=1$ and one recovers the classical growing rate $\beta^* = b_{n^*} C_1^{eq}$. When reactions involving other clusters are possible, this growing rate increases. The mobility of small clusters therefore leads to an increase of the nucleation rate and a decrease of the incubation time by the same factor.

### *Non-Stoichiometric Clusters*

Until now, it has been assumed that clusters have a fixed stoichiometry corresponding to the equilibrium one of the nucleating phase. In some systems, the composition of the nucleating phase can vary. One therefore needs to extend cluster dynamics so as to let the cluster stoichiometry vary [7].
To illustrate such an extension of the formalism, we consider the example of a system where the nucleating phase is composed of two elements, A and B, and we assume that the composition can vary. A cluster is then a group of A and B atoms that are linked by a neighbourhood relation. If the clusters are homogeneous (no segregation of one element at the interface for instance), they can be simply described by two variables: the number $i$ and $j$ of elements A and B they contain. We call $G_{i,j}$ the free energy of such a cluster. If the system is under-saturated, one can show that the concentration of $\{i,j\}$ clusters is given by the distribution

$$C_{i,j}^{eq} = C_0 \exp-\left(-\frac{G_{i,j} - i\mu_A - j\mu_B}{kT}\right), \tag{92}$$

where $\mu_A$ and $\mu_B$ are Lagrange multipliers ensuring matter conservation for A and B and are related to their chemical potentials.

We assume that only monomers are mobile. The time evolution of clusters containing $i$ A elements and $j$ B elements is then governed by the master equation

$$\frac{\partial C_{i,j}}{\partial t} = J_{i-1,j \to i,j} - J_{i,j \to i+1,j} + J_{i,j-1 \to i,j} - J_{i,j \to i,j+1}, \quad \forall \{i,j\} \neq \{1,0\} \text{ and } \{i,j\} \neq \{0,1\},$$

$$\frac{\partial C_{1,0}}{\partial t} = -\sum_{i \geq 0} \sum_{j \geq 0} J_{i,j \to i+1,j} - J_{1,0 \to 2,0}, \tag{93}$$

$$\frac{\partial C_{0,1}}{\partial t} = -\sum_{i \geq 0} \sum_{j \geq 0} J_{i,j \to i,j+1} - J_{0,1 \to 0,2}.$$

Fluxes are written as a difference between the evaporation and the condensation of a monomer:

$$J_{i,j \to i+1,j} = b_{i,j \to i+1,j} C_{1,0} C_{i,j} - \alpha_{i+1,j \to i,j} C_{i+1,j},$$
$$J_{i,j \to i,j+1} = b_{i,j \to i,j+1} C_{0,1} C_{i,j} - \alpha_{i,j+1 \to i,j} C_{i,j+1}. \tag{94}$$

The condensation and evaporation rates are still linked by a detailed balance condition, leading to the relations:

$$\alpha_{i+1,j \to i,j} = b_{i,j \to i+1,j} \exp\left(\frac{G_{i+1,j} - G_{i,j} - G_{1,0}}{kT}\right),$$
$$\alpha_{i,j+1 \to i,j} = b_{i,j \to i,j+1} \exp\left(\frac{G_{i,j+1} - G_{i,j} - G_{0,1}}{kT}\right). \tag{95}$$

One therefore needs a physical modelling of the condensation process so as to express the coefficients $b_{i,j \to i+1,j}$ and $b_{i,j \to i,j+1}$. The evaporation rates are then obtained by Eq. (95) and the kinetics is obtained by integration of Eq. (93).

Starting from the master equation (93), the classical nucleation theory has been extended to treat a multicomponent system. This has been first performed by Reiss [7] for a binary system like the one considered there and then extended by Hirschfelder [8] to a general multicomponent system. Both authors assumed that the growth of the critical nucleus was entirely driven by the free energy. It was realized later by Stauffer [9] that the growth direction in the $\{i, j\}$ plane may also be affected by the condensation coefficients, especially when coefficients corresponding to A and B condensation have very different values. He proposed an expression of the steady-state nucleation rate for the binary system that was then extended to a multicomponent system by Trinkaus [10]. All these approaches calculated the steady-state nucleation rate in the vicinity of the critical nucleus. Wu [11] instead defined a global nucleation rate that should correspond more closely to what can be measured experimentally. In the following, we give the expression of the steady-state nucleation rate for a binary system obeying the master equation (93) in the local approach as derived by Vehkamäki [12]. Expressions in the more general case – multicomponent systems and mobile clusters other than monomers – can be found in her book or in the references quoted above.

The critical cluster corresponds to the saddle point of the cluster formation free energy $\Delta G_{i,j} = G_{i,j} - i\mu_A - j\mu_B$ appearing in the equilibrium distribution (92). It is thus defined by the equations

$$\frac{\partial \Delta G_{i,j}}{\partial i} = 0 \quad \text{and} \quad \frac{\partial \Delta G_{i,j}}{\partial j} = 0. \tag{96}$$

We call $\Delta G^*$ the corresponding formation free energy, and $H^*$ the Hessian matrix calculated for the critical cluster:

$$H^* = \begin{pmatrix} \left.\frac{\partial^2 \Delta G_{i,j}}{\partial i^2}\right|_{\{i,j\}^*} & \left.\frac{\partial^2 \Delta G_{i,j}}{\partial i \partial j}\right|_{\{i,j\}^*} \\ \left.\frac{\partial^2 \Delta G_{i,j}}{\partial i \partial j}\right|_{\{i,j\}^*} & \left.\frac{\partial^2 \Delta G_{i,j}}{\partial j^2}\right|_{\{i,j\}^*} \end{pmatrix}. \tag{97}$$

This Hessian matrix has two eigenvalues. One of them is negative and gives the direction in the $\{i, j\}$ space corresponding to the maximal decrease of the critical cluster free energy. In the approach of Reiss and Hischfelder, this direction corresponds to the nucleation flow. Nevertheless, one should generally take into account that the condensation rates for A and B elements may be different as this will impact the direction of the nucleation flow. We therefore define a new matrix characterizing the condensation process for the critical cluster:

$$B^* = \begin{pmatrix} C_{1,0} b_{i,j \to i+1,j} & 0 \\ 0 & C_{0,1} b_{i,j \to i,j+1} \end{pmatrix}\Bigg|_{\{i,j\}^*}. \tag{98}$$

The fact that this matrix is diagonal reflects our assumption that only reaction involving monomers are possible. The angle $\theta$ of the nucleation flow in the $\{i, j\}$ space is then defined by

$$\tan\theta = \frac{-H_{11}^* B_{11}^* + H_{22}^* B_{22}^* - \sqrt{4H_{12}^{*2} B_{11}^* B_{22}^* + \left(H_{11}^* B_{11}^* - H_{22}^* B_{22}^*\right)^2}}{2H_{12}^* B_{11}^*}. \tag{99}$$

The equivalent of the Zeldovitch factor is given by

$$Z = -\frac{H_{11}^* + 2H_{12}^* \tan\theta + H_{22}^* \tan^2\theta}{\left(1 + \tan^2\theta\right)\sqrt{|\det(H^*)|}}, \tag{100}$$

and the average growth rate of the critical cluster by

$$\beta^* = \frac{\det(B^*)}{B_{11}^* \sin^2\theta + B_{22}^* \cos^2\theta}. \tag{101}$$

With these definitions, the steady-state nucleation rate keeps its usual expression:

$$J^{st} = \beta^* Z C_0 \exp\left(-\frac{\Delta G^*}{kT}\right). \tag{102}$$

*Configurational Frustrations between Clusters*

Cluster dynamics simulations rely on the cluster gas approximation derived in section 0. This thermodynamic approximation, initially introduced by Frenkel [48], is strictly valid only in the dilute limit. It indeed assumes that the space occupied by the clusters can be neglected when computing the configurational partition function (Eq. (45)) of the cluster gas: each cluster only occupies one site whatever its size. Lépinoux [56] has shown that this approximation can be improved so as to properly take into account frustrations between clusters, *i.e.* the space forbidden to a given cluster by other clusters. This allows to model systems which are not as dilute as required by Frenkel's treatment.

We note $V_{j,n}$ the number of sites that a cluster of size $j$ forbids to a cluster of size $n$. According to Lépinoux [56], the equilibrium cluster size distribution is given by

$$C_n^{\text{eq}} = C_0 \exp\left(-\frac{G_n - n\mu}{kT}\right) \exp\left(-\sum_j C_j^{\text{eq}} V_{j,n}\right), \qquad (103)$$

or equivalently

$$C_n^{\text{eq}} = C_0 \left(\frac{C_1^{\text{eq}}}{C_0}\right)^n \exp\left(-\frac{G_n - nG_1}{kT}\right) \exp\left[-\sum_j C_j^{\text{eq}} \left(V_{j,n} - nV_{1,n}\right)\right]. \qquad (104)$$

One clearly sees that Frenkel's approximation corresponds to neglecting all exclusion volumes ($V_{j,n} = 0$). When exclusion volumes are considered, one only obtains an implicit expression of the size distribution: equilibrium cluster size concentrations $C_j^{\text{eq}}$ are required to evaluate the right hand side of Eqs. (103) or (104). A self-consistent loop can be used to evaluate the equilibrium distribution, starting from the distribution given by Frenkel's approximation (Eqs. (49) or (50)).

The exclusion volumes can be approximated by identifying a cluster of size $n$ with a sphere of radius $R_n$. This leads to

$$V_{j,n} = \frac{4\pi}{3}\left(R_j + R_n\right)^3. \qquad (105)$$

The radii $R_n$ depend on the temperature as a cluster becomes less compact with higher temperatures due to its configurational entropy. Nevertheless, one can reasonably assume that these radii are close to the ones corresponding to the more compact cluster shape [56] and use Eq. (61).

The second step is to get the kinetic coefficients $\alpha_n$ and $\beta_n$. As previously, the condensation rate $\beta_n$ is obtained by the proper physical modelling of the condensation process leading to an expression of the form (59). However, it is not possible anymore to assume that the evaporation rate is an intrinsic property of the cluster: the obtained expression would violate the assumption because of the frustration contribution in the cluster size distribution. The constrained equilibrium is not satisfactory either as it leads to a diverging frustration correction, and hence diverging evaporation rates, in supersaturated systems. There is actually no framework that allows deriving rigorously the evaporation rate from the condensation rate taking into account cluster frustrations. It seems that the most reasonable scheme is to consider that the classical expression (64) of the evaporation rate needs to be corrected from frustrations caused by the instantaneous cluster size distribution and not by an hypothetical equilibrium one:

$$\alpha_{n+1}(t) = b_n C_0 \exp\left(\frac{G_{n+1} - G_n - G_1}{kT}\right) \exp\left[\sum_j C_j(t)\left(V_{j,n+1} - V_{j,n} - V_{j,1}\right)\right] \qquad (106)$$

This set of condensation and evaporation rates ensures that the cluster distribution evolve towards the equilibrium distribution given by Eq. (103) for subcritical clusters. Eq. (106) clearly shows that the evaporation rate is not anymore an intrinsic property of the cluster as it now depends on the whole cluster distribution. Moreover, as this parameter depends on the instantaneous concentrations $C_j(t)$, it needs to be calculated at each time step. When the system is dilute, the frustration correction in Eq. (106) becomes negligible and one recovers the classical expression of the condensation rate. Comparisons with atomic simulations [50, 56] have shown that this treatment of cluster frustrations greatly improves the ability of cluster dynamics to describe nucleation kinetics for high supersaturations.

### *Limitations of the Cluster Description*

The above extensions of cluster dynamics have allowed removing two limitations of classical nucleation theories due to initial simplifying assumptions:
- Only monomers are mobile and therefore only reactions involving monomers are possible.
- The cluster stoichiometry is fixed and known *a priori*. It is assumed to correspond to the composition of the nucleating phase at equilibrium with the mother phase.

The extension to mobile clusters is quite straightforward, and that to non stoichiometric clusters shows that it was possible to take into account a non-fixed cluster composition. The composition of the nucleating cluster was found as the one minimizing the work $\Delta G^*$ necessary to form them.

  Nevertheless, some limitations still remain for this nucleation modelling approach. One of these limitations arises from the needed assumption that clusters are homogeneous. This assumption is induced by the fact that clusters are only described by the number of elements they contain. This is not always valid as segregation may occur in some systems: it can be more favourable for one element to lie at the interface between the cluster and the matrix instead of in the core of the cluster. In such a case, one needs to introduce at least one more parameter so as to describe the cluster structure. Binder and Stauffer [4] have extended cluster dynamics formalism so as to incorporate additional parameters describing cluster internal degrees of freedom, but the application of the formalism appears then quite intricate.

  Cahn and Hilliard [60] proposed a modelling approach different from the classical one presented here which is based on a cluster description. Their approach agrees with the classical one at low supersaturations and underlines some limitations of the classical approach with increasing supersaturations. They showed that the work $\Delta G^*$ required to form a critical cluster becomes progressively less than that given by the classical theory and approaches continuously to zero at the spinodal limit, thus for a finite supersaturation. By contrast, the classical theory predicts that this work becomes zero only for an infinite supersaturation (Eq. (6)). Moreover, the classical theory assumes that clusters are homogeneous and that their composition is the one minimizing the work $\Delta G^*$. Cahn and Hilliard showed that the composition at the centre of the nucleus approaches that of the exterior mother phase when the supersaturation tends to the spinodal limit, and that the interface becomes more diffuse until eventually no part of the nucleus is even approximately homogeneous. The last disagreement found with the classical theory is the variation of the critical cluster size. They showed that this size first decreases, passes through a minimum and then increases to become infinite when the supersaturation increases and approaches the spinodal limit. Nevertheless, some recent experiment observations [61, 62] have contradicted this last point, showing no divergence of the cluster critical size when approaching the spinodal limit.

# Conclusions

  Two different approaches based on an equivalent cluster description can therefore be used to model nucleation in a phase separating system. In the classical nucleation theory, one gets expressions of the nucleation rate and of the incubation time. These expressions depend on a limited number of input parameters: the nucleation driving force, the interface free energy and the condensation rate. On the other hand, the kinetic description of nucleation relies on a master equation. Cluster dynamics simulations, *i.e.* the integration of this master equation, allow then to obtain the time evolution of the cluster size distribution. The input parameters needed by such simulations are the cluster condensation rates and the cluster free energies. At variance with classical nucleation theory, no external thermodynamic model is needed to calculate the nucleation driving force: cluster dynamics simulations possess their own thermodynamic model, the cluster gas. As shown above, both approaches are intrinsically linked. But it is worth saying that they differ in the way they can be used to model the kinetics of phase transformations. Classical nucleation theory is able to model only the nucleation stage. To model the whole kinetics, one needs to couple this theory with classical descriptions of the growth and coarsening stage. Such a coupling can be done following Wagner and Kampmann approach [63, 64]. On the other hand, the cluster dynamics modelling approach is not restricted to the nucleation stage. It as well predicts the growth and the coarsening kinetics.

To conclude, one should say that this cluster approach is well suited when one knows what the nucleating new phase looks like. Such information is not always available a priori. One has then to use other modelling techniques. These can be atomic simulations, like molecular dynamics [65, 66], for condensation of a gas into a liquid or crystallisation of a liquid, or kinetic Monte Carlo [67, 67, 69] for solid-solid phase transformations, or phase field simulations (see Appendix). These simulations are computationally much more time consuming and, as a consequence, are limited to the study of high enough supersaturations. Nevertheless, they can be very useful to understand what happens in the nucleation stage so as to build then a classical model based on a cluster description and extend the range of supersaturations that can be simulated. Moreover, these atomic or phase field simulations can be a convenient way to calculate the input parameters needed by classical theories.

## Appendix: Phase Field Simulations

The phase field approach describes the different phases through continuous fields like the atomic concentration or long-range order parameters. The spatial and temporal evolution of the microstructure is then driven by differential equations obeyed by these fields. As this technique is the object of the article Phase-field Modeling of Microstructure Evolution in this Volume, this Appendix addresses how nucleation can be handled in such simulations.

The main advantage of phase field simulations is that all spatial information on the microstructure is obtained. This is to contrast with classical approaches where a limited number of information is known, like the flux of nucleating particles (classical nucleation theory) or the cluster size distribution (cluster dynamics). This may make the phase field approach an attractive technique for modeling nucleation in specific situations. Such simulations indeed perfectly take into account phase inhomogeneities. These inhomogeneities can be, for instance, a solute segregation in the vicinity of a defect like a dislocation. Phase field simulations therefore allow describing heterogeneous nucleation associated to a variation of the driving force. Moreover, in the case of solid-solid phase transformations, the elastic energy is fully contained in the calculation of the system free energy [41]. One therefore does not need a specific expression for the elastic self energy of a nucleating particle nor for its elastic interaction with the existing microstructure. The correlated and collective nucleation due to elastic interaction between precipitates is naturally described. Two different roads have been proposed to include nucleation in phase field simulations.

One can use the phase field approach to calculate spatial variations of the concentrations and of the order parameters describing the different phases as well as the inhomogeneous strain created by the microstructure. One then calculate the nucleation free energy as a function of the local phase fields and of the local strain. Finally the expression of the nucleation rate given by classical theory is used to seed the phase field simulations with new nuclei [70, 71, 72]. In this way, one obtains a spatial variation of the nucleation rate caused by the microstructure inhomogeneities.

The phase field approach offers another way to model nucleation without relying on the classical theory. One can add to the equations describing the phase field evolution a stochastic term through a Langevin force so as to describe thermal fluctuations. This will allow nucleation to proceed. Phase field simulations can then naturally describe the spatial and temporal evolution of the microstructure, from the nucleation to the coarsening stage [73, 74, 75]. Nevertheless, this description is usually only qualitative: to obtain a fully quantitative modeling, the amplitude of the Langevin force needs to be carefully set. In particular, it has to depend on the coarse-graining size like the other ingredients of the simulation (chemical potentials, mobilities, stiffness coefficients) [76]. Such phase field simulations which naturally handle nucleation through thermal fluctuations suffer from the small time-step needed to catch the rare event of a nucleating particle. On the other hand, simulations using an explicit description of the nucleation do not have this drawback.

# Acknowledgments

The author thanks Bernard Legrand, Georges Martin, Maylise Nastar, and Frédéric Soisson for fruitful discussions and their careful reading of the manuscript.

# Selected References

### *References Focusing on Nucleation*

### *More General References on Phase Transformations*